\documentclass[twocolumn, deluxetables]{aastex62}
\usepackage{amsmath}
\usepackage[caption=false]{subfig}

\graphicspath{{./}{figures/}}

\shorttitle{The Disk Wind in GRO J1655-40}
\shortauthors{Balakrishnan et al.}

\newcommand{\gro}{GRO J1655-40}

\begin{document}

\title{Swift Spectroscopy of the Accretion Disk Wind in the Black Hole \gro}

\correspondingauthor{M.~Balakrishnan}
\email{bmayura@umich.edu}

\author{M.~Balakrishnan}
\author{J. M. Miller}
\author{N.~Trueba}
\author{M.~Reynolds}
\affil{Department of Astronomy, The Univ. of Michigan, 1085 S. University Ave., Ann Arbor, MI, 48109, USA}
\author{J. Raymond}
\affil{Harvard-Smithsonian Center for Astrophysics, 60 Garden Street, Cambridge, MA, 02138, USA}
\author{D. Proga}
\affil{Department of Physics, University of Nevada, Las Vegas, Las Vegas, NV, 89154, USA }
\author{A. C. Fabian}
\affil{Institute of Astronomy, University of Cambridge, Madingley Road, Cambridge 3B3 OHA, UK}
\author{T. Kallman}
\affil{NASA Goddard Space Flight Center, Code 662, Greenbelt, MD, 20771, USA}
\author{J. Kaastra}
\affil{SRON Netherlands Institute for Space Research, Sorbonnelaan 2, 3584 Utrecht, The Netherlands}

\begin{abstract}
{\it Chandra} obtained two High Energy Transmission Grating (HETG) spectra of the stellar-mass black hole GRO J1655$-$40 during its 2005 outburst, revealing a rich and complex disk wind.  Soon after its launch, the {\it Neil Gehrels Swift Observatory} began monitoring the same outburst.  Some X-ray Telescope (XRT) observations were obtained in a mode that makes it impossible to remove strong Mn calibration lines, so the Fe K$\alpha$ line region in the spectra was previously neglected.  However, these lines enable a precise calibration of the energy scale, facilitating studies of the absorption-dominated disk wind and its velocity shifts.  Here, we present fits to 15 {\it Swift}/XRT
spectra, revealing variability and evolution in the outflow.  The data strongly point to a magnetically driven disk wind: both the higher velocity (e.g., $v \simeq 10^{4}~{\rm km}~{\rm s}^{-1}$) and lower velocity (e.g., $v \simeq 10^{3}~{\rm km}~{\rm s}^{-1}$) wind components are typically much faster than is possible for thermally driven outflows ($v\leq 200~{\rm km}~{\rm s}^{-1}$), and photoionization modeling yields absorption radii that are two orders of magnitude below the Compton radius that defines the typical inner extent of thermal winds.  Moreover, correlations between key wind parameters yield an average absorption measure distribution (AMD) that is consistent with magnetohydrodynamic wind models.  We discuss our results in terms of recent observational
and theoretical studies of black hole accretion disks and outflows, and future prospects.
\end{abstract}


\section{Introduction} \label{sec:intro}
Stellar-mass black holes are a window into many intriguing and
fundamental astrophysical questions, including explorations of general
relativity, the origin and demographics of black hole spin, the
relationship between supernovae and gamma-ray bursts, basic accretion
disk physics, and the physics of wind outflows and relativistic jets.
Many of the most interesting phenomena observed in stellar-mass black
holes are only manifested at certain fractions of the Eddington
luminosity, likely signaling that "states" -- periods of coupled
multi-wavelength behavior -- represent major changes to the geometry
and processes within the accretion flow (for a review, see, e.g.,
\citealt{2006ARA&A..44...49R}).  It is notable, for instance, that
strong winds (most clearly seen in ionized X-ray absorption lines) and
relativistic jets (imaged in radio bands) are generally not present at
the same time \citep{2006ApJ...646..394M, 2008ApJ...680.1359M, 2009Natur.458..481N, 2012ApJ...746L..20K} .  This can potentially be explained by, e.g., changes in the the dominant magnetic field geometry of the accretion disk with Eddington fraction \citep{2015ApJ...809..118B}.

Thermal continuum emission from the accretion disk is prominent in the
X-ray spectra of stellar-mass black holes at high Eddington fractions.
Indeed, the ``thermal--dominant'' state is defined by strong disk
emission, and very low rms variability (see, e.g., \cite{2006ARA&A..44...49R}). Magnetic viscosity in the form of magneto-rotational instabilities is the only plausible way to drive accretion in these disks \citep[e.g.][]{1991ApJ...376..214B}, but it
leaves no specific imprint on continuum spectra.  

Winds from the inner accretion disk are very likely powered by magnetic processes and can potentially put limits on the magnetic field that emerges from the disk. These winds test \citep[e.g.][]{2015ApJ...814...87M, 2016ApJ...821L...9M} the fundamental assumptions of the $\alpha$-disk model introduced by \citealt{1973IAUS...55..155S}. Accretion disk winds can also be launched from the outer disk via thermal driving, and this process could be important in creating the central X-ray corona and regulating the overall mass accretion rate through the disk \citep{1973IAUS...55..155S,1983ApJ...271...70B, 1986ApJ...306...90S}. However, new numerical simulations suggest that the regulating effects of a wind are likely to be more modest \citep{2020ApJ...890...54G}.

Some winds may expel more gas than is able to accrete onto the black hole. However, a number of uncertainties in the launching radius, density, and filling factor complicate estimates of the mass outflow rate and kinetic power of winds, apart from a few exceptional cases \citep[see][]{2012ApJ...746L..20K, 2015ApJ...813L..37K, 2015ApJ...814...87M, 2016ApJ...821L...9M}; for a
review of winds, see \citealt{2012MNRAS.422L..11P}.  Extremely sensitive,
high-resolution spectroscopy is one means of attempting to measure the
crucial details that enable robust wind properties and driving mechanisms; the study of wind evolution represents another.  Unfortunately, the latter avenue has been largely inaccessible, as practical considerations make it difficult for {\it Chandra} to observe a single outburst more than a few times.  Moreover, the modes of operation that enable CCD spectrometers to observe bright Galactic transients typically suffer from effects (e.g., charge transfer inefficiency) that lead to large systematic errors in velocity shifts.

GRO J1655$-$40 is a well-known and particularly well-studied
stellar-mass black hole.  It was discovered during an outburst in July 1994 \citep{1995Natur.374..703H} and was notable for the detection of apparently superluminal radio jets \citep{1995Natur.374..141T}.  The relative strength of the approaching and receding jets suggested that the inner disk in GRO J1655$-$40 could be viewed at an angle as high as $\theta = 85^{\circ}$, while optical observations constrain the inclination to $\theta \approx 70^{\circ}$ \citep{1997ApJ...477..876O, 2001ApJ...554.1290G}. After this outburst, the source was quiescent until early 1996.  Optical studies of the binary during this quiescence led to stringent constraints on the masses and distance of the binary.  Orosz \& Bailyn measured a black hole mass of $M_{BH} = 7.0\pm 0.2~M_{\odot}$, a companion mass of of $M_{C} = 2.3\pm 0.1~M_{\odot}$, and a source distance of 3.2~kpc \citep{1997ApJ...477..876O}.  These measurements enable accurate estimates of the source luminosity and Eddington fraction (see Figure 1).

A second outburst in 1997 was extensively studied with {\it RXTE},
revealing the evolution of the continuum emission, broadband X-ray
variability properties \citep[see][]{1999ApJ...520..776S}, and notably
the detection of high frequency QPOs at 300 and 450~Hz \citep{2001ApJ...552L..49S}.  After a long quiescent period, the source was again observed in outburst starting in February, 2005 \citep{2005ATel..414....1M}.  During this outburst, {\it Chandra} obtained two high-resolution spectra in bright states with prominent disk emission.

In one of these exposures, a particularly complex disk wind was observed.
Fe XXII lines were detected that constrained the
density to be approximately $n = 10^{14}~{\rm cm}^{-3}$.  This
enabled a direct radius constraint through the ionization parameter
formalism, $\xi = L/nr^{2}$ (where $L$ is the luminosity, $n$ is the
hydrogen number density, and $r$ is the radius at which the absorption
is observed).  Small absorption radii, approximately $r \simeq
500-1000~GM/c^{2}$, were indicated.  Such radius estimates are effectively upper limits, since the launching radius must be smaller than the absorption radius, and the more ionized parts of the disk wind (traced by He-like Fe XXV and H-like Fe XXVI) likely originate even closer to the black hole.  Plausible thermal winds can only be launched from radii about two orders of magnitude larger, suggesting that the wind in GRO J1655$-$40 is magnetically driven.  

A study of the average Absorption Measure Distribution (AMD), which probes the relationship between hydrogen column density and ionisation parameter \citep[see][]{2007ApJ...663..799H}, in this spectrum has recently found that its structure matches predictions for magnetic winds \citep{2017NatAs...1E..62F}.

Although {\it Chandra}/HETG spectroscopy is powerful, GRO J1655$-$40 was only observed using this configuration on two occasions (March 12th and April 1st, 2005), as seen in Figure 1.  Other black
hole outbursts have been observed on four or fewer occasions.  These
observations have sampled different source states, finding that winds
are generally anti-correlated with jets
\citep[see][]{2006Natur.441..953M, 2008ApJ...680.1359M, 2009Natur.458..481N, 2012ApJ...746L..20K, 2012MNRAS.422L..11P}.  However, the {\em evolution} of an accretion flow is
not effectively sampled by, e.g., two observations in disk--dominated
states where winds are found.  Herein, we present an analysis of 15
{\it Swift}/XRT \citep{2004ApJ...611.1005G} spectra of GRO~J1655$-$40 obtained during its 2005 outburst (for a prior treatment, see Brocksopp et al.\ 2006).  The wind is detected in each
of these spectra.  Section 2 details the instrumental set-up and our
reduction of the data.  The spectral analysis we have undertaken is
described in Section 3.  In Section 4, we discuss our findings in the
context of other studies of black hole accretion.

\section{Data Reduction}
We analyzed 15 {\it Swift}/XRT observations of GRO J1655$-$40,
obtained between March 19 and May 22, 2005, and available in the
public HEASARC archive.  The start time, duration, and identification
number of each observation are given in Table 1.  The timing of these observations can be seen in Figure 1, in addition to the RXTE flux and hardness ratios at these times. Each of the {\it Swift}
observations was obtained in ``photodiode'' (PD) mode.  This is a fast
timing mode achieved by alternately performing one serial clock shift
and one parallel clock shift. The result is a very rapid clocking of
each pixel across any given point on the CCD.  PD mode provides 0.14
ms timing resolution, but no spatial information.  The XRT
automatically switches between Piled-up Photodiode (PUPD) mode and
low-rate photodiode (LrPD) mode, depending on the source brightness.
In PUPD mode, all pixels are captured, whereas in LrPD mode, only
photons below a certain threshold are captured.

\begin{figure}[ht]
\centering
\includegraphics[width=0.45\textwidth]{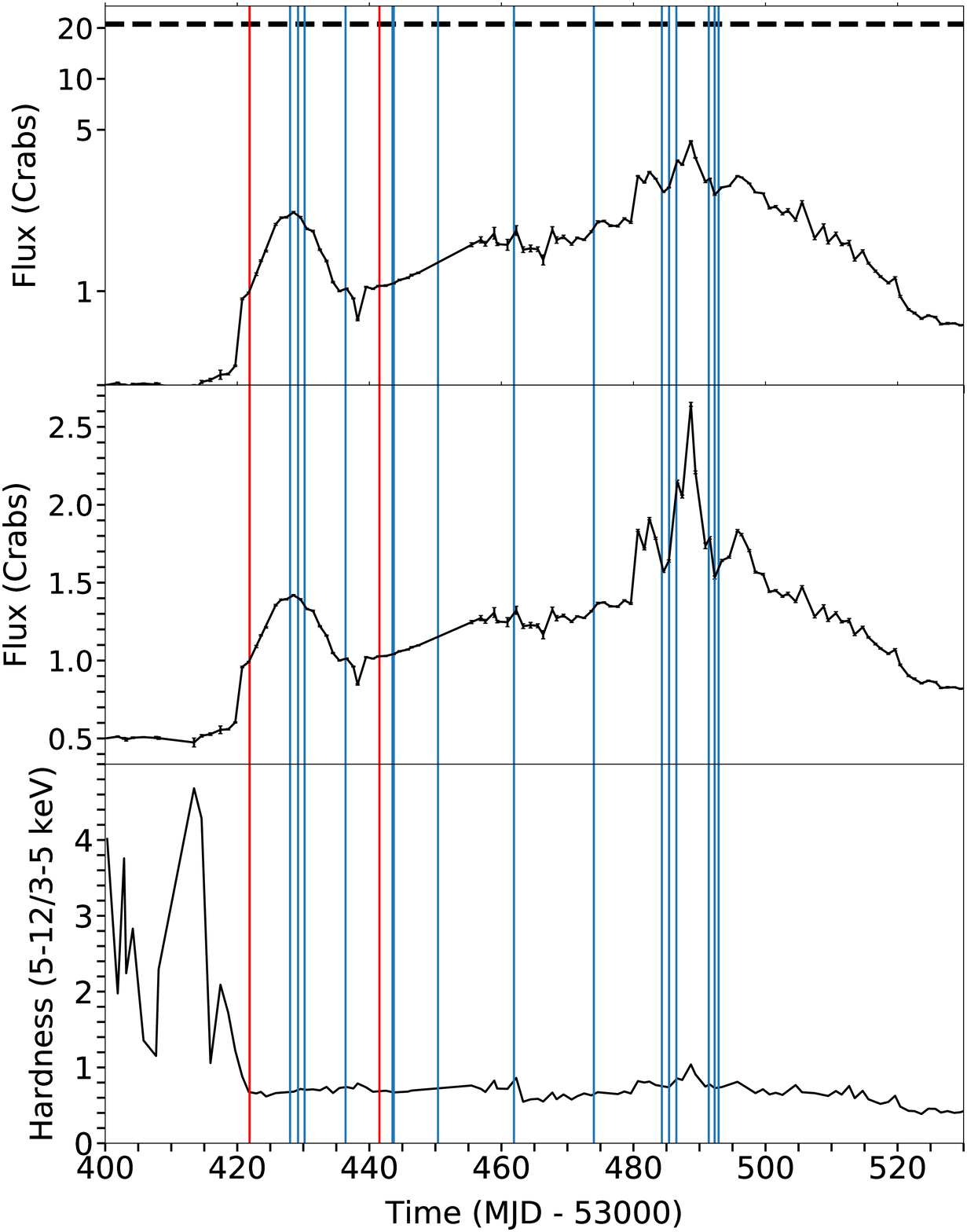}
\caption{The RXTE All-Sky Monitor (ASM) light curve and hardness ratio from a portion of the 2005 outburst of GRO~J1655$-$40.  The top two panels show the 1.5--12.0~keV light curve, converted to Crab units assuming that 1 Crab is 70 counts per second. The first panel plots the flux in log scale, with a dotted black line corresponding to the Eddington luminosity. The second plot shows the flux with a linear scale. The hardness ratio (5-12 keV/3-5 keV) is shown in the bottom panel.  Values for these curves come from one-day averaged data from the ASM.  The blue vertical lines represent the dates of the 15 {\it Swift}/XRT observations used in our analysis, while the two red lines show the dates of the two {\it Chandra}/HETGS observations that were taken during this outburst.}
\end{figure}

We processed these observations the using tools available in HEASOFT
version 6.26.  Initial event cleaning was carried out using the task
\texttt{XRTPIPELINE}. Spectra were extracted using \texttt{XSELECT}.
No regions were used as the whole CCD is exposed as a unit; for the
same reason, we were unable to separately extract background regions.
At the flux observed from GRO J1655$-$40, however, the background is
negligible.  The spectra were later binned to a S/N ratio of 10 using
\texttt{ftgrouppha}.  Since regions cannot be used, custom responses
are not required, and we used the default redistribution matrix file (RMF) and ancillary response file (ARF) for this mode (available in the calibration data base).

It must be noted that a micrometeoroid struck the XRT CCD in 2005
\citep{2006MNRAS.365.1203B}.  The hot pixels and other damage
caused by the impact made it impossible to operate the XRT in PD mode
after the strike.  Very few sources were observed in PD mode, so few
publications have resulted, and the community and mission have not
refined its calibration in the way that other modes have been refined
over 15 years of mission operation.

\section{Analysis} 
Spectral fits were made using XSPEC version 12.10.1 \citep{1996ASPC..101...17A}.  In
all cases, a lower bound of 1~keV was adopted because calibration
uncertainties appear to be stronger at lower energy.  In most cases,
we adopted an upper fitting bound of 9~keV, owing to calibration
uncertainties and/or modest photon pile-up at higher energy.  Strong
pile-up is only expected at a flux above 3~Crab.  In a few specific cases, a
different upper bound was adopted as unphysical fit residuals were observed
(e.g., an increasing trend in the data/model ratio above 9 keV).  The
fits minimized the $\chi^{2}$ fit statistic, and all errors reported in this work reflect 1 $\sigma$ errors calculated in \texttt{XSPEC}, to maintain consistency. The errors on derived quantities were calculated using standard error propagation, and are also 1 $\sigma$ confidence errors.

Following many studies of black hole spectral evolution, we adopted a simple continuum model,
\texttt{tbabs}$\times$ (\texttt{diskbb} $+$ \texttt{power-law}).  The
\texttt{tbabs} component describes absorption in the interstellar
medium \citep{2000ApJ...542..914W}, and its single parameter, the equivalent hydrogen
column density $N_{H}$, was allowed to vary.  The \texttt{diskbb}
component is the familiar multi-color disk model \citep{1984PASJ...36..741M}, characterized by the disk color temperature ($kT$) and its
flux normalization.  Both parameters were allowed to vary without
bounds.  Finally, the \texttt{power-law} component is a simple
power-law function, characterized by a photon index ($\Gamma$) and
flux normalization.  We set a lower bound of $\Gamma = 1.0$ on the power-law index.

On its own, this continuum does not allow for a good fit.  In the
non-imaging PD mode, the Mn calibration lines that are created by the $^{55}$Fe radioactive source -- Mn K$\alpha_{1}$ and Mn
K$\alpha_{2}$ (5.889 and 5.899~keV), and Mn~K$\beta$ (6.49~keV) remain
in the spectrum and dominate in the Fe K band.  We fit these lines
using simple Gaussian functions, fixed at their laboratory
wavelengths, with linked widths ($\sigma$) and variable flux
normalizations.  The widths are consistent with zero at the resolution
of the CCD.  The lines are so strong that they {\it must} be modeled
well to have a good overall fit (see Figure 2).  It is clear that the observed line
energies are not consistent with their laboratory values, owing to a
shift in the detector response in PD mode.  We therefore modified our
entire spectral model by the function \texttt{zmshift}, which shifts
the model as if it were affected by a bulk velocity shift (e.g., a cosmological red-shift).  
Typically, we measure values of $z = -0.01$, and then a good continuum
fit is achieved.  

This method of calibrating the energy scale is illustrated in Figure 2.  Observation 00030009010 is shown, fit with the model described in Tables 2 and 3.  The
strong, narrow Mn calibration lines are the most prominent discrete features in
the spectrum.  In the left-hand panel, the best-fit model is shown
after setting the overall shift to zero.  In the middle panel, the
model is shown at its best-fit value of $z = -0.010$.  In the right-hand panel, the ionized absorption is then also modeled.  When the Mn calibration lines are modeled properly, the
energy scale is correctly determined, at least in the region close to the lines.  The $v=0.01c$ blue-shifts that we later measure in wind absorption lines are therefore real, and evaluated relative to the proper zero-point owing to the action of \texttt{zmshift}.

\begin{deluxetable*}{lrrl}
\tablewidth{700pt}
\tabletypesize{\scriptsize}
\tablecaption{Swift/XRT observation details.}
\tablehead{
\colhead{\hspace{.01cm}OBSID\hspace{.01cm}} & \colhead{MJD} & 
\colhead{Exposure Time (ks)} & \colhead{Start Date}
}
\startdata
00030009005 & 53448   & 1.709                                                        & 2005-03-19 \\
00030009006 & 53449.2 & 1.580                                                        & 2005-03-20 \\
00030009007 & 53450.2 & 1.437                                                        & 2005-03-21 \\
00030009008 & 53456.4 & 2.838                                                        & 2005-03-27 \\
00030009009 & 53463.5 & 2.395                                                        & 2005-04-03 \\
00030009010 & 53463.7 & 2.455                                                        & 2005-04-03 \\
00030009011 & 53470.4 & 2.025                                                        & 2005-04-10 \\
00030009012 & 53481.9 & 0.214                                                        & 2005-04-21 \\
00030009014 & 53494   & 1.139                                                        & 2005-05-03 \\
00030009015 & 53504.3 & 1.999                                                        & 2005-05-14 \\
00030009016 & 53505.4 & 0.909                                                        & 2005-05-15 \\
00030009017 & 53506.5 & 1.945                                                        & 2005-05-16 \\
00030009018 & 53511.4 & 1.909                                                        & 2005-05-21 \\
00030009019 & 53512.3 & 2.345                                                        & 2005-05-22 \\
00030009020 & 53512.9 & 0.725                                                        & 2005-05-22
\enddata
\end{deluxetable*}

\begin{figure*}[ht]
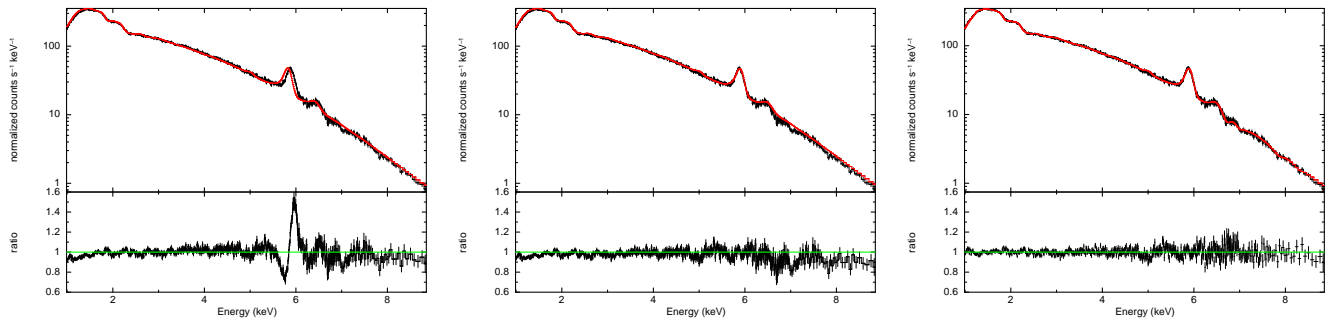

\centering
	   \includegraphics[angle=270,width=0.325\textwidth]{fig2a.eps}
	   \includegraphics[angle=270,width=0.325\textwidth]{fig2b.eps}
	   \includegraphics[angle=270,width=0.325\textwidth]{fig2c.eps}
\caption{Fits to the {\it Swift}/XRT spectrum of GRO J1655$-$40 (observation 00030009010). In the left panel, the value of \texttt{zmshift} has been set to zero and the ionised absorption model has been removed from the fit. In the middle, the value of \texttt{zmshift} has been set to the best fit value and the ionised absorption model has been removed from the fit. In the right panel, the value of \texttt{zmshift} has been set to the best fit value and the ionised absorption is fully modeled. The shift is determined utilizing the known, laboratory-determined energy of the Mn calibration lines.
}
\end{figure*}

Our ultimate goal is to model Fe K absorption lines, at energies only
slightly above the the Mn calibration lines.  As a check on the
efficacy of \texttt{zmshift} and the calibration lines to correctly
determine the energy scale, Figure 3 compares the Fe K absorption
lines seen in {\it Swift} observation 00030009010 (obtained on 2005 April 03) to
those seen in the {\it Chandra} observation of GRO J1655$-$40 that was obtained only
two days prior (2005 April 01).  Within Figure 3, the energy scale
of the {\it Swift}/XRT spectrum has been shifted by the value measured
with \texttt{zmshift}; the energy scale of the {\it Chandra}/HEG
spectrum is unaltered.  The lines are deeper in the HEG spectrum owing
to its higher resolution; the question is whether the line centroid
energies agree.  The He-like Fe XXV and Fe XXVI lines match perfectly.
The blue-shift of the Fe XXVI line core is readily apparent in the XRT
spectrum, and so too is the blue wing on this line that trends toward
even higher blue-shifts.  The Fe XXV He$\beta$ line is detected in
the XRT, and so too are lines from rare elements between 5.5--6.3~keV (Cr and Mn).
Overall, the degree of correspondence is excellent, and this check
validates the ability of the calibration lines and our modeling
procedure to measure accurate wind properties.

\subsection{Photoionized wind models} 

As seen in Figure 3, the {\it Swift}/XRT spectra of GRO J1655$-$40
detect multiple absorption lines, closely matching those found in the contemporaneous {\it Chandra}/HEG spectrum.   We therefore
used the same grid of XSTAR photoionization models employed to fit the {\it Chandra} spectra in \citealt{2008ApJ...680.1359M} and \citealt{2015ApJ...814...87M}.  This particular grid has
high energy resolution, solar elemental abundances, and suitable spacing in both $\xi$ and $N_{H}$.  A number density of $n = 10^{14}~{\rm cm}^{-2}$ was used to construct the grid, based on Fe XXII lines in the {\it
  Chandra} spectrum (later verified based on re-emission from the
wind; see \citealt{2015ApJ...814...87M}).  The assumed input spectrum is the
continuum observed in the {\it Chandra} observation; however, the
spectra obtained with {\it Swift} are broadly similar.  The XSTAR grid has an upper limit column density of $N_{H} = 6\times 10^{23}~{\rm cm}^{-2}$ and an upper limit ionization of log$\xi = 6$. We note that XSTAR does not correct for Thomson scattering; we estimate that this could lead to an under-estimate of the luminosity by a maximum of 40\% at the very highest column density values.  This uncertainty is comparable to the uncertainty in luminosity that is incurred with any choice of a specific continuum model. 

\begin{figure*}[ht]
\centering
\includegraphics[width=0.99\textwidth]{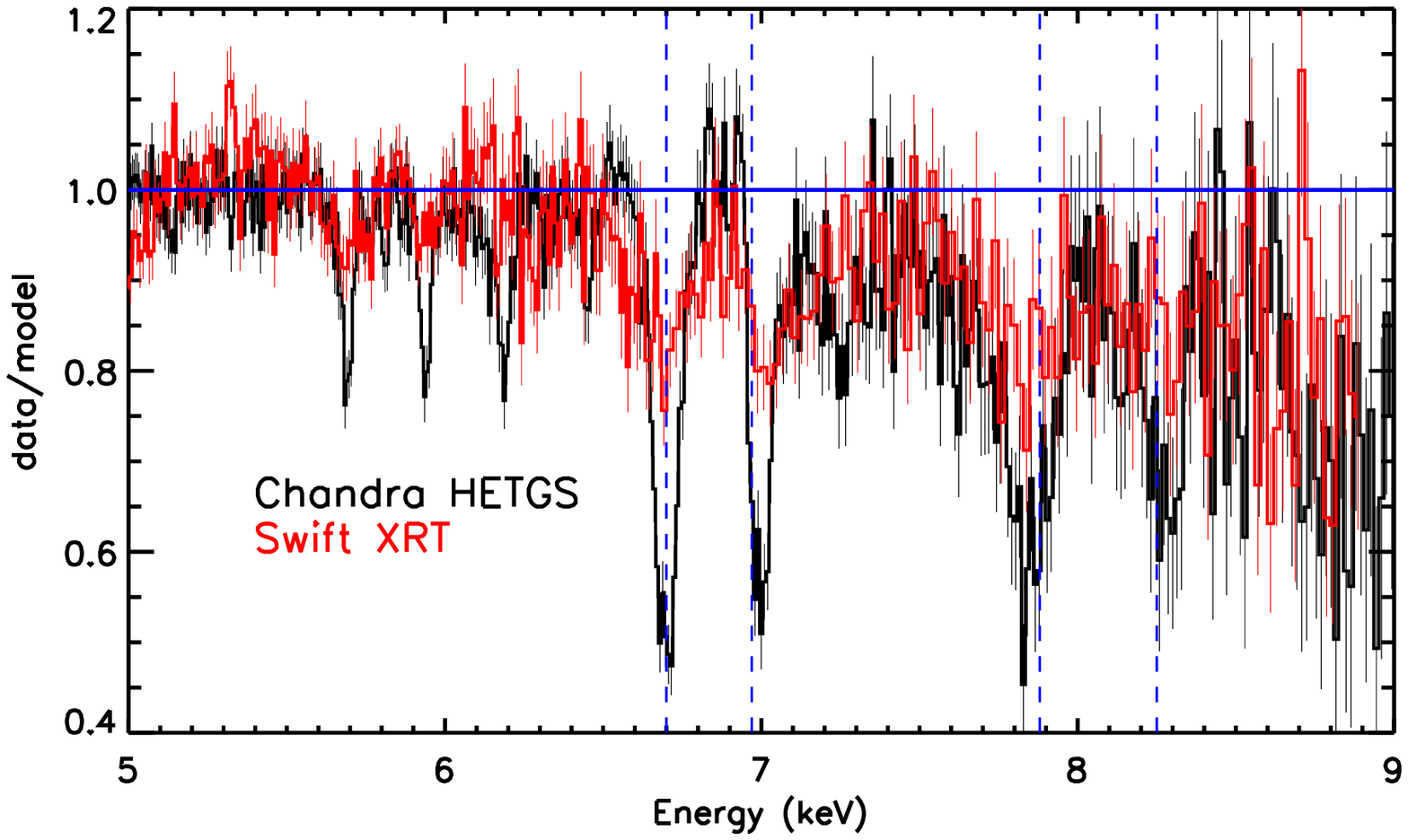}
\caption{A comparison of contemporaneous {\it Chandra}/HETG and {\it Swift}/XRT spectra of GRO J1655$-$40.  Both spectra are binned for visual clarity.  The {\it Chandra} spectrum was obtained on 2005 April 01; the {\it Swift} spectrum was obtained on 2005 April 03 (observation 00030009010).  The energy scale of the {\it Swift} spectrum was corrected utilizing the Mn calibration lines present in PD mode (see Figure 2).  The resulting agreement of the spectra is excellent, and the XRT spectrum is sensitive enough to even detect He-like Cr XXIII at 5.68~keV. The blue vertical lines indicate the laboratory wavelengths of the He-like Fe XXV and H-like Fe XXVI resonance lines.  The stronger blue-shift of the Fe XXVI line clearly indicates that there are multiple velocity components within the disk wind.}
\end{figure*}

\begin{figure*}[ht]
\centering
\includegraphics[width=0.95\textwidth]{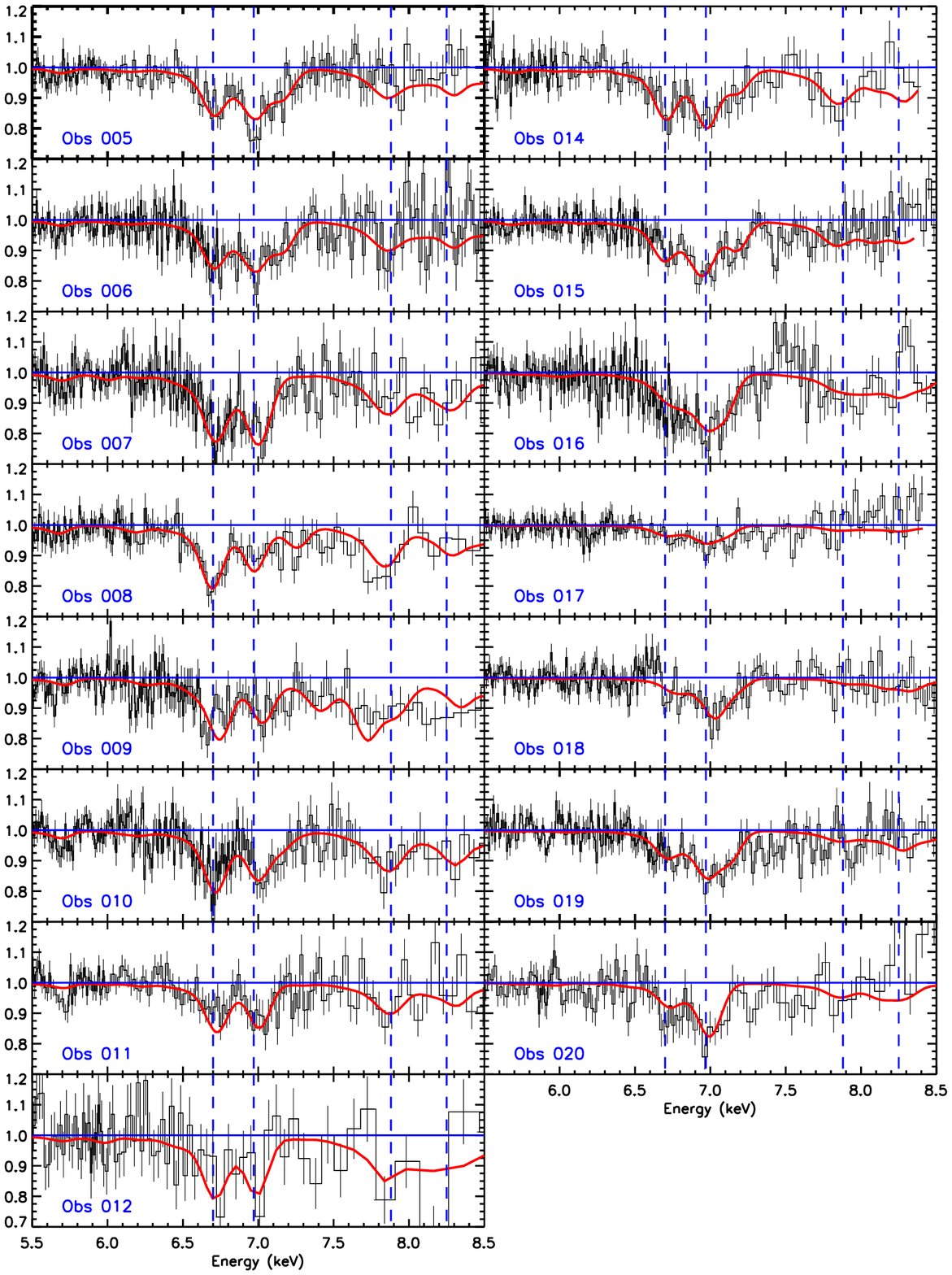}
\vspace{-0.5in}
\caption{{\it Swift}/XRT spectra in the Fe~K band (black), and best-fit models (red).  The energy scale was established through fits to strong Mn K calibration lines, and shifted to reflect the calibration.  The spectra are shown as a ratio to the best-fit continuum model in order to focus on the Fe K absorption lines over the Mn K calibration lines.  The model for each spectrum includes two XSTAR photoionization zones.  Dashed, vertical lines indicate the laboratory wavelengths of the Fe He$\alpha$ and Ly$\alpha$ (6.70~keV and 6.97~keV) and He$\beta$ and Ly$\beta$ lines (7.88~keV and 8.25~keV).  Many of the Fe Ly$\alpha$ lines show stronger blue-shifts than the corresponding Fe He$\alpha$ lines, and sometimes extended blue wings, indicating multiple velocity components and strong shifts in the disk wind.}
\end{figure*}

For consistency, we fit all of the spectra with two photionization zones with fully independent parameters (we define "Zone 1" as the higher-velocity zone in all cases, and "Zone 2" as the lower- velocity zone).  The measured
parameters for each zone include the equivalent neutral hydrogen
column density ($N_{H}$), the log of the ionization parameter (${\rm
  log}\xi$), and the velocity shift ($v/c$).  
  Our final model, then,
is as follows: \texttt{zmshift}$\times$(\texttt{Gauss}$_{1}$ + \texttt{Gauss}$_{2}$
$+$ \texttt{Gauss}$_{3}$ $+$ \texttt{tbabs}$\times$
\texttt{XSTAR}$_{1}$ $\times$ \texttt{XSTAR}$_{2} \times$ (\texttt{diskbb}$+$ \texttt{power-law})).

\section{Results}

Figure 4 shows each of the 15 {\it Swift}/XRT spectra, and the best-fit model with two photoionization zones.  The spectra are shown as a ratio to the best-fit continuum, so that the wind absorption lines can be examined in detail.  The ratio is particularly useful in that it is directly related to the equivalent width of the wind lines, and therefore to the column density in the wind.  It is evident that the column density in the wind varies significantly.  The Fe XXV He$\alpha$ and Fe XXVI Ly$\alpha$ lines are very clearly detected; in a few spectra, the He$\beta$ and Ly$\beta$ lines are also detected.  In all cases, the blue-shift of the absorption is apparent.  Indeed, the Fe XXVI Ly$\alpha$ line is typically blue-shifted even more strongly than the Fe XXV He$\alpha$ line, clearly indicating the need for multiple velocity zones.  In many cases, the Fe XXVI Ly$\alpha$ line has a particularly extended blue wing, indicating strong blue-shifts reaching up to $v\simeq 0.01-0.03c$.  

Table 2 lists the continuum fit parameters that were measured in our analysis, as well as the 
observed and unabsorbed flux in the 1--9~keV band (see Figure 5).  The value of the fit statistic and the degrees of freedom for the {\it total} spectral model, including zero, one, or two absorption zones, is also given in Table 2.  The disk temperature and normalizations reported in Table 2 are the most robust continuum parameters: the soft part of the 1--9~keV fitting range is not subject to strong photon pile-up effects, and it also covers the peak of the disk blackbody distribution for the measured range of temperatures.  It is also the case that the disk component strongly dominates every spectrum in the 1--9~keV band.  The equivalent neutral hydrogen column density is measured to vary between $4.5 \leq N_{H} / 10^{21}~{\rm cm}^{-2} \leq 7.2$.  This unexpected variability is likely driven by calibration and response issues, not by variations in a low-ionization component of the wind (which would lead to variations above the expected baseline value of $N_{H} = 7.4\times 10^{21}~{\rm cm}^{-2}$, \citealt{1990ARA&A..28..215D}); for this reason, the column density is not listed in Table 2.  Most of the power-law indices are unphysically flat, especially for a soft, disk--dominated state.  This may be partly instrumental, and partly due to the effects of photon pile-up.  It is clear that the power-law component is only dominant at the high energy end of the 1--9~keV band, however, and it has only a very minor effect on the observed and derived wind properties.

Table 3 lists the properties of the wind that are directly measured from the spectrum, via the grid of XSTAR models.  To facilitate comparisons, the parameter values for each of the two zones are listed separately.  The ionization parameter formalism implies that the wind can potentially originate at small radii with non-zero gravitational red-shift, so Table 3 lists corrected velocity values (measured relative to the red-shift at the launching radius).  The measured column density, ionization, launching radius, and outflow velocity of each zone are plotted versus time in Figure 6.  

Table 4 lists the Compton radius, and derived values of the launching radii and photoionization radii as defined below.
We define the photoionization radius as $r_{\rm phot} = L / N_H \xi$; this is an upper limit since it assumes a filling factor of $f=1$.   We define the launching radius as $r_{\rm launch} = \sqrt{L / n \xi}$, assuming $n = 10^{14}~{\rm cm}^{-3}$ \citep{2008ApJ...680.1359M}.  This radius is also plotted in Figure 6 and listed in Table 4, as a formal quantity with errors.  However, $r_{\rm launch}$ as defined here is really also an upper limit, because the true launching point of the wind is likely interior to the point at which the density-sensitive Fe XXII lines were produced, and realistic wind density profiles fall with radius. The approximate agreement of the approximate agreement of $r_{\rm launch}$ and $r_{\rm phot}$ suggests that the filling factor is not very small ($f \approx 0.3$).

In the same panel in Figure 6, the Compton radius, $R_{C}$, and $R = 0.1\times R_{C}$ are also plotted (in red and blue, respectively).  These mark the smallest radii at which thermal winds can be launched in seminal theoretical treatments of such winds \citep{1983ApJ...271...70B, 1996ApJ...461..767W, 2010ApJ...719..515L} .  The Compton radius is given by 

\begin{equation*}
R_{C} = \frac{10^{10}}{T_{C,8}}\times \frac{M}{M_{\odot}}~{\rm cm}. 
\end{equation*}

\noindent We have assumed that the Compton temperature, $T_{\rm C,8}$, is equal to the characteristic disk temperature in each spectrum, which is true unless the power-law component dominates the spectrum.  This is not the case in the disk--dominated high/soft state wherein winds are detected.  The Compton radius is also tabulated in Table 4.  The characteristic wind radii are all an order of magnitude lower than even $0.1\times R_{C}$, and therefore two orders of magnitude smaller than $R_{C}$.  

The velocity of the faster wind zone is regularly in the vicinity of $v \simeq 10^{4}~{\rm km}~{\rm s}^{-1}$, or $v/c \simeq 0.03$.  This exceeds even the fastest component previously reported in the richest {\it Chandra} \citep{2015ApJ...814...87M} spectrum.  Where good measurements are possible, the slower wind zone is typically consistent with a velocity of $v \simeq 10^{3}~{\rm km}~{\rm s}^{-1}$, significantly higher than predicted in theoretical simulations of thermal winds ($v \leq 200~{\rm km}~{\rm s}^{-1}$, \citealt{2017ApJ...836...42H}).  However, it is important to note that 60\% of the velocities in Zone 2 are broadly consistent with zero. This is the result of modest resolution inhibiting the detection of small velocity shifts in relatively weak lines.  

Table 5 lists the key wind feedback quantities that can be derived from the directly measured quantities: the mass outflow rate, $\dot{M}_{\rm wind}$, and the kinetic power $L_{\rm kin}$.  These parameters were calculated using the following equations:

\begin{equation*}
    \dot{M}_{\rm wind} = \Omega f \mu m_p v \frac{L_{ion}}{\xi}
\end{equation*}

\begin{equation*}
    L_{\rm kin} = 0.5 \ \dot{M}_{\rm wind} v^{2}
\end{equation*}

\noindent where $\Omega$ is the geometric covering factor (we assumed $\Omega = 1.5$ in our XSTAR grids), $\mu$ is the mean atomic weight ($\mu = 1.23$ was assumed), $m_{p}$ is the mass of the proton, $v$ is the (redshift-corrected) outflow velocity, $L_{ion}$ is the luminosity measured in the 1--9~keV band, and $\xi$ is the measured ionization parameter.  In our calculations, the volume filling factor, $f$ is assumed to be unity.  This is an upper limit, and it is very likely that the value is lower.  We have opted not to calculate and tabulate a single filling factor because the density that we have assumed is likely an under-estimate (as noted previously), and because there are two zones that may have distinct filling factors.  Figure 7 plots the mass outflow rate and kinetic power versus time, as derived quantities with two-sided errors.  Applying any filling factor will simply slide the distributions along the vertical axis within these plots, and so the relative trends are likely reliable. It is possible that the filling factor varies with time, but we expect that any such variations are small, and this is bolstered by the fact that fits to the $N_{H}$ versus $r_{\rm launch}$ and $N_{H}$ versus $\xi$ relationships give consistent results. The most important result to be drawn from Table 5 and Figure 7 is that the mass outflow rate and kinetic power are variable, by at least an order of magnitude.
Some of the values plotted in Figure 7 carry large errors owing to the growth of errors in the propagation.

The AMD characterizes the distribution of the hydrogen column density along the line of sight as a function of ${\rm log}\xi$ and is defined in \citealt{2007ApJ...663..799H} as $\rm AMD = d N_H / d ({\rm log}\xi)$. This can be re-cast \citep[see][]{2019ApJ...886..104T} into expressions of the form $n \propto r^{-\alpha}$, where $\alpha$ is a simple index, not the magnetic stress term from \citealt{1973IAUS...55..155S}.  A spherically symmetric, mass-conserving flow would have a value of $\alpha = 2$, for instance.  In contrast, if the wind is distributed over large scales, as per a magnetically driven wind, values of $1.0 \leq \alpha \leq 1.5$ are expected (\citealt{2009ApJ...703.1346B, 2016A&A...589A.119C, 2017NatAs...1E..62F}; we note that these treatments do not define $\alpha$ in a consistent manner).  

Figure 8 plots the log of the column density, versus the log of the launching radius.  (Recall that in this analysis, the launching radius is a quantity with errors given by $r_{\rm launch} = \sqrt{L/n\xi}$ and assuming a density of $n = 10^{14}~{\rm cm}^{-3}$, whereas the photoionization radius can be regarded as an upper limit and is likely larger than the launching radius).  These parameters are found to be strongly anticorrelated, with a p-value of $p \leq 1.1\times 10^{-5}$.  It must be noted that the value of alpha does not depend on the assumed density, since a different value of the density would yield a different normalization, not a different slope. Figure 9 plots the log of the column density versus the log of the ionization parameter.  The index was calculated using $\alpha = \frac{2-2x}{2-x}$ (Figure 8) and $\alpha = \frac{1+2x}{1+x}$ (Figure 9); these expressions come from our derivations shown in Appendix A.  Values within the range $1.28 \leq \alpha \leq 1.54$ are consistent with all of the relationships that we examined.

\begin{figure}[ht]
\centering
\includegraphics[width=0.45\textwidth]{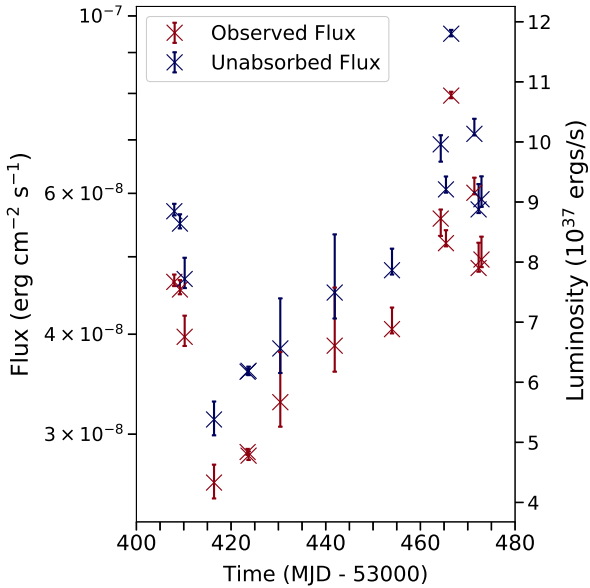}
\caption{Observed flux and unabsorbed flux for all observations over the 1-9 keV range. The unabsorbed flux was measured setting the column density of the internal obscuration to zero, and calculating the flux.}
\end{figure}

At high ionization parameters, Fe XXVI dominates, and it can be difficult to obtain strong limits from above.  The column density can then also run to high values, since it is possible to model with a high column without creating a large number of strong lines.  We there examined the possibility that this effect could distort our measurement of the AMD.  First, The \texttt{steppar} command in \texttt{XSPEC} was run to construct 160x120 grids in $N_{H}$ and log($\xi$).  In general, these ran from $1 \leq N_{\rm H}/10^{22}~{\rm cm}^{-2} \leq 60$ and $4 \leq {\rm log}(\xi
)\leq 6$.  We flagged zones where the area enclosed by the $2\sigma$ confidence region was greater than 10\% of the parameter space, or where this contour overlapped the upper ionization limit of our grid.  A small number of zones are excluded using this criteria.  Though there are fewer total points in the relationships used to derive values of $\alpha$ when these restrictions are enforced, the values of $\alpha$ that are derived are fully consistent with those found when all zones are considered (see Figure 8 and Figure 9).  We therefore conclude that the relatively high ionization levels encountered in X-ray binaries has not distorted our view of the AMD in GRO J1655$-$40.

\begin{figure*}[ht]
\centering
\includegraphics[height=0.9\textheight]{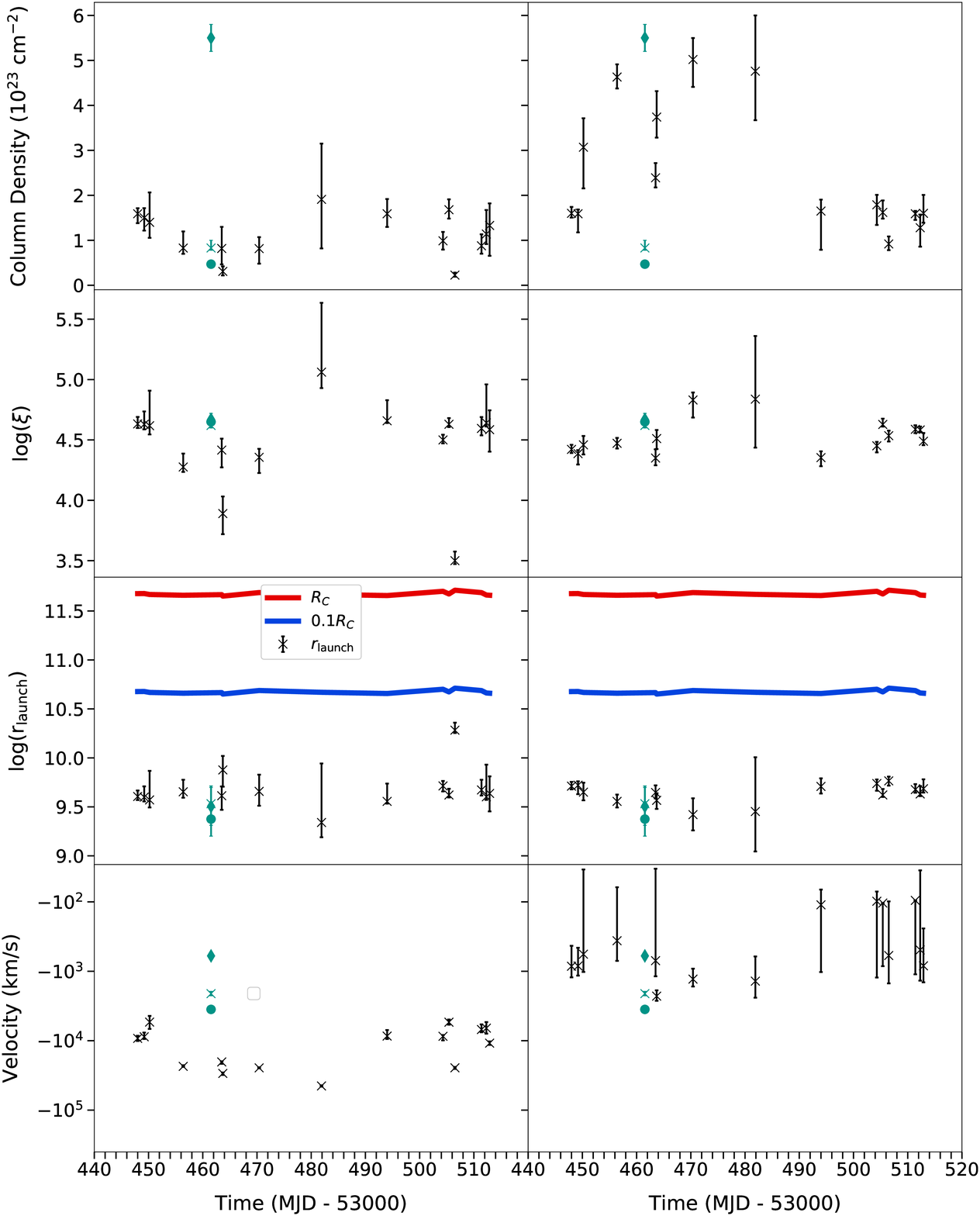}
\vspace {-0.5in}
\caption{Values for column density, ionisation parameter, launching radius, and velocity are presented. The fits yield two absorption zones per observation, which are represented in the side-by-side plots (left: Zone 1; right: Zone 2). The green points in the panels correspond to the values presented in \citealt{2015ApJ...814...87M}. The column density and ionisation parameter change over the course of the outburst. The ionization in both zones is generally quite high, and broadly comparable, apart from two observations that have relatively low ionizations in Zone 1 ($\rm \log{\xi} = 3.89$ and $3.5$.) The launching radius also varies over time, and is shown in log units.  For comparison, we plot $0.1 R_C$ and $R_C$ in blue and red, respectively, to illustrate the smallest and most likely launching radius for Compton-heated thermal winds.  In the velocity panel, we show velocities that have been corrected for the gravitational redshift at $r_{\rm launch}$.  The highest blueshifts are very well constrained. }
\end{figure*}

\begin{figure}[ht]
\centering
\includegraphics[width=0.45\textwidth]{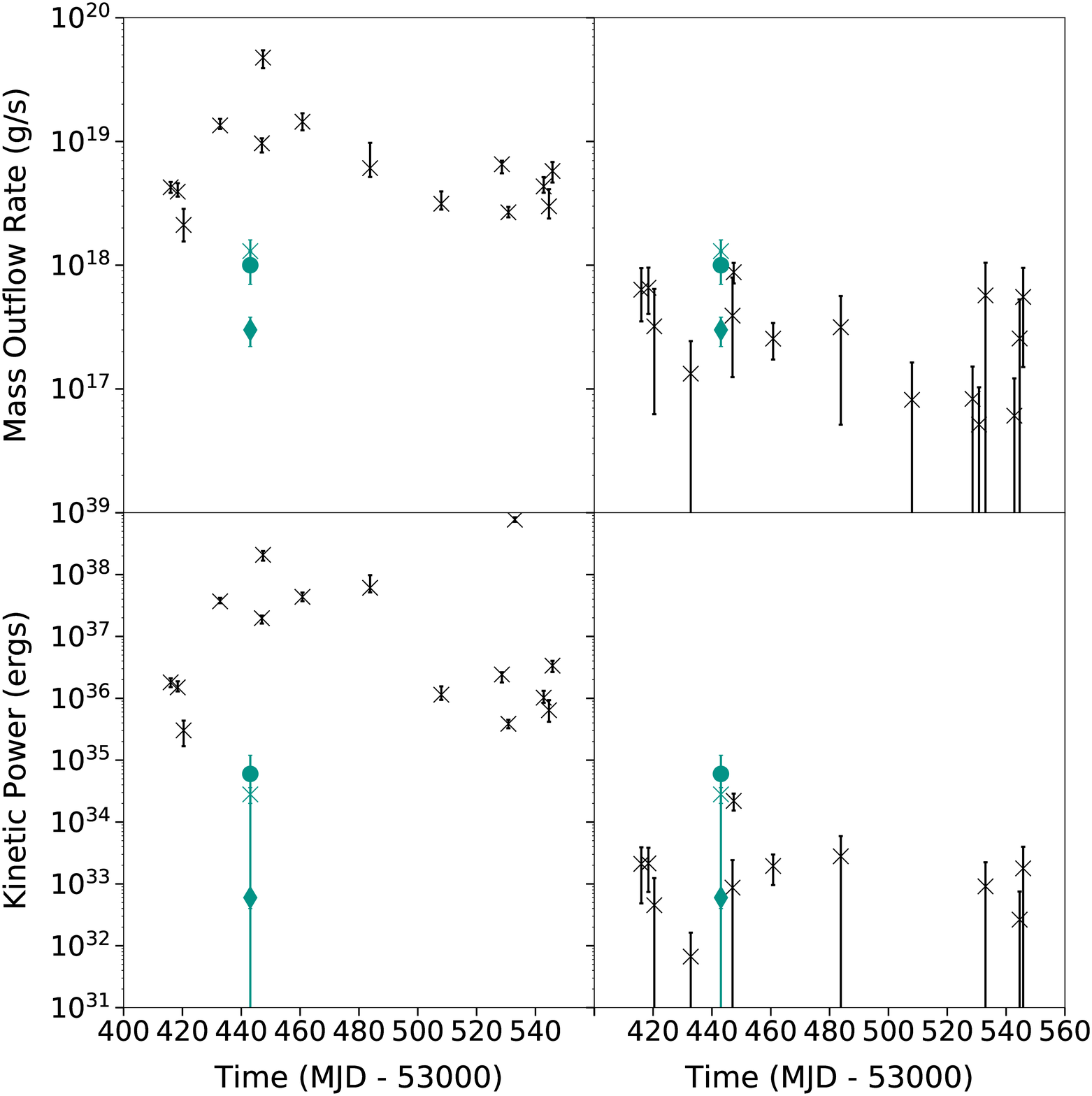}
\caption{Mass outflow rate (g/s) and kinetic power (ergs) for all wind components are shown (left: Zone 1; right: Zone 2).  The error bars are very large on values for Zone 2, owing to the propagation of uncertainties on the velocity.  The values presented in each of these plots should be regarded as upper limits, as they do not account for a filling factor, $f$.}
\end{figure}

\section{Discussion}
We analyzed 15 {\it Swift}/XRT spectra of the stellar-mass black hole GRO J1655$-$40, obtained during its 2005 outburst.  The spectra were obtained in "photodiode" mode, which is no longer available owing to a micrometeoroid strike on the detector.  By using the Mn K calibration lines intrinsic to this mode, we obtained an excellent calibration of the energy scale in the Fe K band and carefully studied the nature and evolution of the rich accretion disk wind.  Although modest in an absolute sense, these 15 disk wind spectra represent an important increase in the number of detections obtained from a single transient outburst, and therefore the chance to study disk winds in a new way.  A number of important results are obtained, including the detection of clear variability in critical wind parameters, persistently high velocities and small launching radii, and constraints on the AMD of the wind.  In this section, we summarize the wind properties that are measured from the spectra, examine the most likely wind driving mechanisms, and scrutinize some of the limitations of our  work.

Figure 6 clearly shows that plausible launching radii for the disk wind are consistently at least order of magnitude smaller than the innermost possible launching radius for thermal winds ($0.1 R_{c}$, where $R_{C}$ is the Compton radius), and two orders of magnitude lower than likely thermal wind launching radii, provided a low Eddington fraction.  The true innner launching point of the wind is likely smaller than that probed by the parameter we have called $r_{launch}$ because this is calculated assuming that $n = 10^{14}~{\rm cm}^{-3}$ based on fits to Fe XXII lines in the {\it Chandra} spectra of GRO J1655$-$40.  The photoionization radius, $r_{phot}$, is calculated without assuming any value for the density.  It is a strict upper limit in that it assumes a filling factor of unity, and its values range between $164$ $GM/c^{2}$ to $2.81 \times 10^4 $ $GM/c^{2}$.  These upper limits are still lower than 0.1 $R_{C}$, which varies between $4.14 \times 10^4 $ $GM/c^{2}$ - $4.96 \times 10^4$ $GM/c^{2}$ .  In a small number of cases, $r_{phot} < r_{launch}$, which can only be true if the average density is sometimes higher than $n = 10^{14}~{\rm cm}^{-3}$.

The velocity of the faster wind zone, ($v \gtrsim 10^{4}~ {\rm km}~{\rm s}^{-1}$) is persistently two orders of magnitude above the highest velocities predicted in advanced thermal wind simulations ($v \leq 200~ {\rm km}~{\rm s}^{-1}$, \citealt{2017ApJ...836...42H}).  At CCD resolution, it is more difficult to constrain the velocity of the slower wind component, but those zones that confidently exclude zero shift have best-fit values close to $v = 10^{3}~ {\rm km}~ {\rm s}^{-1}$.  Given that the wind is highly ionized and that radiation pressure on lines is therefore ineffective, only magnetic driving is plausible.  

The results thaw we have obtained with {\it Swift}, including small launching radii and the need for magnetic driving, broadly confirm prior studies of the {\it Chandra} spectra of GRO J1655$-$40 \citep[e.g.][]{2008ApJ...680.1359M, 2015ApJ...814...87M, 2009ApJ...701..865K, 2012ApJ...750...27N}.  The fact of 15 observations allows us to put the richest {\it Chandra} wind spectrum into context.  In Figure 6, the green points correspond to the wind parameters measured by {\it Chandra} and reported in Miller et al.\ (2015).  Those values are within the scatter of the wind parameter values measured with {\it Swift}.  The clear implication is that the {\it Chandra} observation did not catch a rare or ephemeral disk wind or state, but instead sampled a wind with properties that endured over at least 70 calendar days and a factor of three in source flux (also see Figure 1).

The fact of 15 spectra also enabled us to explore the average AMD of the variable disk wind, providing an independent angle on its driving mechanism.  We find that values of $\alpha$ in the range $1.28 \leq \alpha \leq 1.54$ are consistent between two relationships and both absorption zones. It is important to note that the slopes do not depend on the assumed density of the wind.  The derived values are formally consistent with the results of simulations of magnetic winds undertaken by \citealt{2016A&A...589A.119C} that found $\alpha = 1.4$, and with the value of $\alpha = 1.2$ measured in the richest {\it Chandra} spectrum of GRO J1655$-$40 \citep{2017NatAs...1E..62F}.  

It should be noted that \citealt{2016A&A...589A.119C} and \citealt{2017NatAs...1E..62F} used self-similar models, while simulations of magneto-thermal disk winds that do not assume self-similarity show that poloidal fields can actually suppress thermal winds instead of launching them \citep{2018MNRAS.481.2628W}.  This may indicate that the observed winds are magnetohydrodynamic winds driven by the magnetorotational instability within the disk, rather than magnetocentrifugal winds that depend on rigid field lines.

The goal of measuring the AMD in an AGN spectrum is to understand its wind structure (e.g, Behar 2009).  However, we have relied on a series of observations rather than a single spectrum.  Does the AMD that we have constructed really achieve the same end and retain its intended meaning?  

The longer timescales inherent in AGNs mean that wind structure can be studied on dynamical time scales.  This is not possible in stellar-mass black holes: every current X-ray spectrum, and also the AMD measured in GRO J1655$-$40 by Fukumura et al.\ (2017), is an average that samples over numerous dynamical time scales in the wind ($t_{dyn} \sim r/v$, or about 1-10s for the wind in GRO J1655$-$40).  Whereas an AMD in an AGN can capture a snapshot of particular wind conditions, an AMD from a single spectrum of a stellar-mass black hole necessarily captures average or typical wind conditions.  The latter is arguably more effective at understanding the typical structure of a wind, and therefore also its driving mechanisms.

The variability that is sampled by examining numerous observations can further help to reveal the structure of winds in a stellar-mass black hole.  
Changes in the ionizing flux will shift the innermost radius at which a wind spread over much of the disk (e.g., a magnetic wind) can be detected using X-ray absorption lines.  In this way, flux changes can serve to reveal the run of wind properties with radius.  However, a wind that is only launched from large radii (e.g., a Compton-heated thermal wind) will not be affected by flux variations, rather only by changes in the peak of the spectrum.  In our analysis, the disk temperature was taken as an estimate of the Compton temperature.  We find that the photoionization radius of the wind in GRO J1655$-$40 changes by an order of magnitude, while the peak disk temperature only varies by $\Delta (kT) \simeq \pm20\%$.  This more simplistic test also points to a magnetic wind that likely spans a range of radii; the AMD that we have constructed merely echoes this result.   Thus, although the mechanics are somewhat different, our adaptation of the AMD is likely to be an effective diagnostic of the wind structure, consistent with the purpose of the AMD analyses undertaken in single spectra of AGN.

It has been suggested that the wind of GRO J1655$-$40 is extreme, or even unique, and that its differences from other winds reflect that it is driven through enhanced radiation pressure in a Compton-thick, super-Eddington accretion flow (e.g., Tomaru et al.\ 2018).  The wind properties of three other sub-Eddington black holes studied in Miller et al.\ (2015) are broadly similar, and a recent study of 4U 1630$-$472 by Trueba et al.\ (2019) finds a number of detailed similarities.  It is clear, then, that the wind in GRO J1655$-$40 is not unique, but is it super-Eddington?  

The projected outflow velocities that we have measured  are orders of magnitude higher than predicted for such enhanced thermal winds (see the application of results from \citealt{2018MNRAS.481.2628W} in \citealt{2018MNRAS.473..838D}).   However, they are an order of magnitude below the velocities observed in many sources that are clearly super-Eddington.   The X-ray lines in SS 433 are observed to have shifts of $v \simeq 0.2c$, consistent with velocities observed in other wavelengths \citep[e.g.][]{2002ApJ...564..941M}.  The outflows implied in a growing number of ultra-luminous X-ray sources have a similar velocity \citep[e.g.][]{2016Natur.533...64P}.  The disk wind observed in the quasar PDS 456 with {\it XMM-Newton} and {\it NuSTAR} is even more extreme, with a velocity of $v = 0.3c$ \citep{2015Sci...347..860N}.  

It is also clear in Figure 1 that the observed flux is an order of magnitude below that corresponding to the Eddington flux for GRO J1655$-$40.  Indeed, the alternative model requires that the source be Compton-thick, transmitting little of the flux from the central engine.  This invariably results in a strong reflection spectrum from illuminated material out of the direct line of sight; it is this reflection spectrum that defines Seyfert-2 AGN in X-rays.  Indeed, in Compton-thick AGN, the Fe~K$\alpha$ emission line is observed to be several times stronger than the local continuum \citep[e.g.][]{2019ApJ...877..102K}.  This is not confined to AGN: the outflow observed during a super-Eddington outburst in the stellar-mass black hole V404 Cyg was not as fast as many super-Eddington sources, but it manifested extremely strong {\it emission} lines above an implausibly weak continuum, indicative of a Compton-thick flow that effectively blocked the central engine from view \citep[e.g.][]{2015ApJ...813L..37K}.  The spectrum and velocities observed from GRO J1655$-$40 do not match observations of known Compton-thick and/or super-Eddington sources.  If a geometry and/or set of conditions manifested in GRO J1655$-$40 that inhibited the production and/or detection of strong reflection in a Compton-thick, super-Eddington accretion flow, it must be different than the flows seen in other X-ray binaries and AGN.

\citealt{2016ApJ...822...20N} suggested that the infrared flux observed from GRO J1655$-$40 could be understood as an optically thick shell resulting from a super-Eddington episode.  However, this is implausible as it would require a shell of gas to be optically thick in infrared but optically thin in X-rays.  We are unaware of any setting wherein this has been observed.  The monitoring observations obtained with {\it Swift} would require this geometry to hold for at least 70 days, which is inconsistent with a short-lived super-Eddington phase ejecting a shell.

There are some important limitations to our analysis.  We note that most of the spectra are fit well, but none of the fits are formally acceptable ($\chi^{2}/\nu = 1.00$).  This is likely due to modest photon pile-up, and remaining calibration uncertainties, especially where the effective area changes rapidly in the 1.5--2.5~keV band.  The calibration of the energy scale that is enabled by the onboard Mn K lines is valid in the Fe K band, but effects such as charge transfer inefficiency (CTI) are energy-dependent and therefore the calibration may not be as good at low energy.  Since ``photodiode'' mode is no longer operable, and few observations were obtained in this mode, it is unlikely that its calibration will be refined in the future.  

It is also the case that at least one spectrum may require a more complex model.  The spectrum obtained in observation 00030009012 appears to be marginally discrepant with the others.  In this case, issues with the fit are likely astrophysical rather than instrumental.   The Fe XXV He$\alpha$ absorption line appears to be slightly red-shifted (see Figure 4).  The spectrum is marginally better fit with three zones, where one is allowed to be red-shifted.  It is possible that this component marks a ``failed'' wind component, but it is more likely that it represents the red-shift of a (temporarily) stationary disk atmosphere (see Trueba et al.\ 2020).  An observed Fe XXV line energy of approximately 6.65~keV represents a shift of $z\simeq 0.0075$, corresponding to a radius of $ r \simeq 270~GM/c^{2}$.  This is within the launching radius inferred using the density prescribed by the Fe XXII lines, but potentially consistent with the innermost extent of the wind given that most of the flow is more highly ionized and likely has a higher density.  

This analysis gives a glimpse of the discovery space that is opened by obtaining many sensitive, calibrated spectra of a stellar-mass black hole at moderate resolution.  It implies an even greater potential for high-cadence monitoring at high spectral resolution.  So, this glimpse raises the question: can at least 15 excellent disk wind spectra be obtained from a single outburst, either now or in the near future?  Or, is 15 spectra a limit that will not be exceeded?  

The critical element of such observations is the ability to clearly distinguish outflows from static disk atmospheres; this is partly a matter of spectral resolution, partly a matter of sensitivity, and partly a matter of calibration and related systematic error margins.  Astrophysical features that could potentially be used as local standards of rest may not be effective substitutes: the "neutral" Fe~K$\alpha$ line can actually represent a run of charge states between Fe I-XVII with rest energies of 6.40-6.43~keV -- a range of $\Delta v \simeq 1400~{\rm km}~ {\rm s}^{-1}$.  Definitive studies of disk winds in X-ray binaries will likely be possible with {\it XRISM} \citep{2018SPIE10699E..22T} if its operations can accommodate high-cadence monitoring.  

We thank the anonymous referee for comments and suggestions that improved this manuscript.  We gratefully acknowledge helpful conversations with Jamie Kennea.

\begin{figure*}[ht]
\centering
\includegraphics[width=0.49\textwidth]{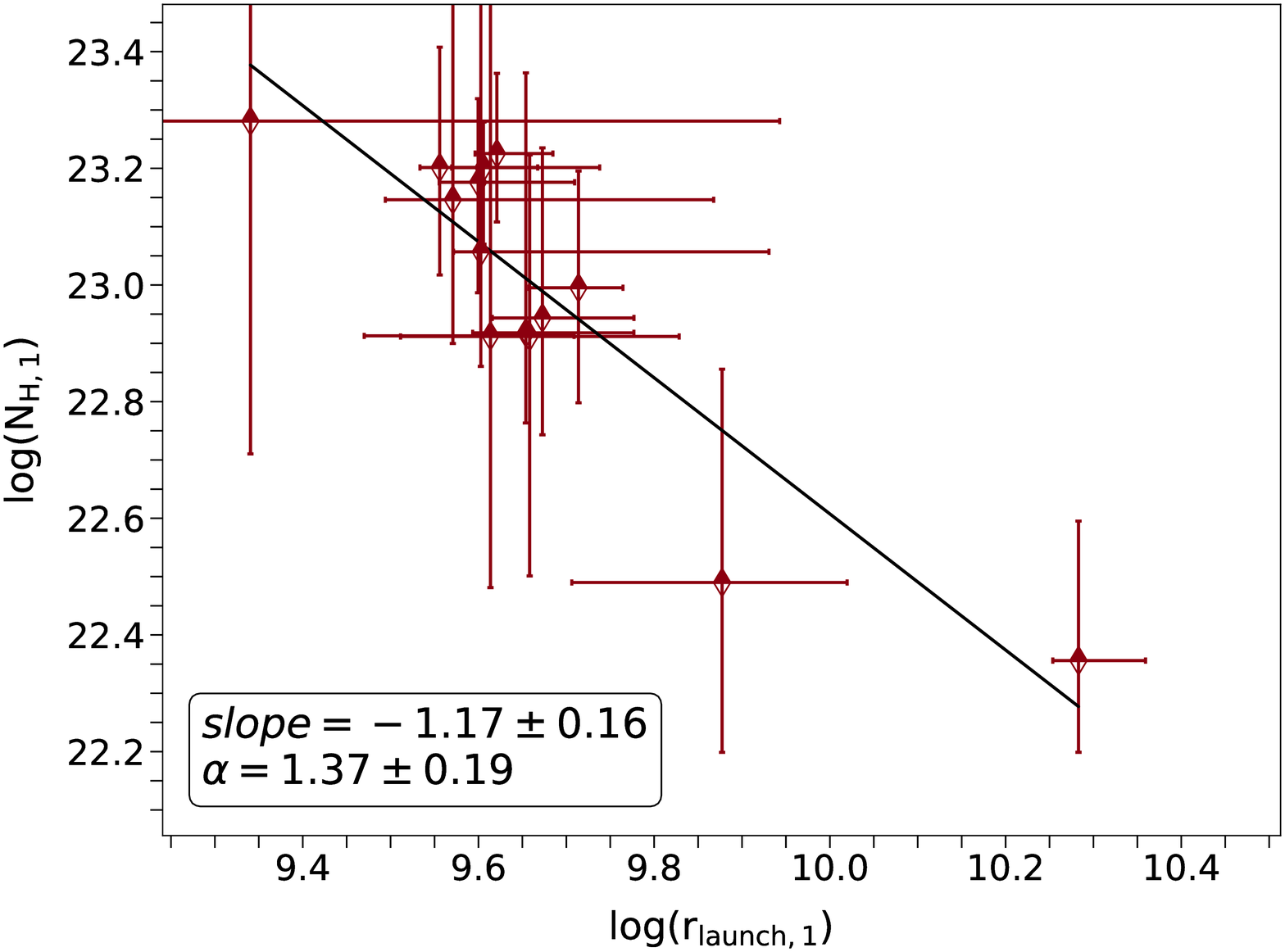}
\includegraphics[width=0.49\textwidth]{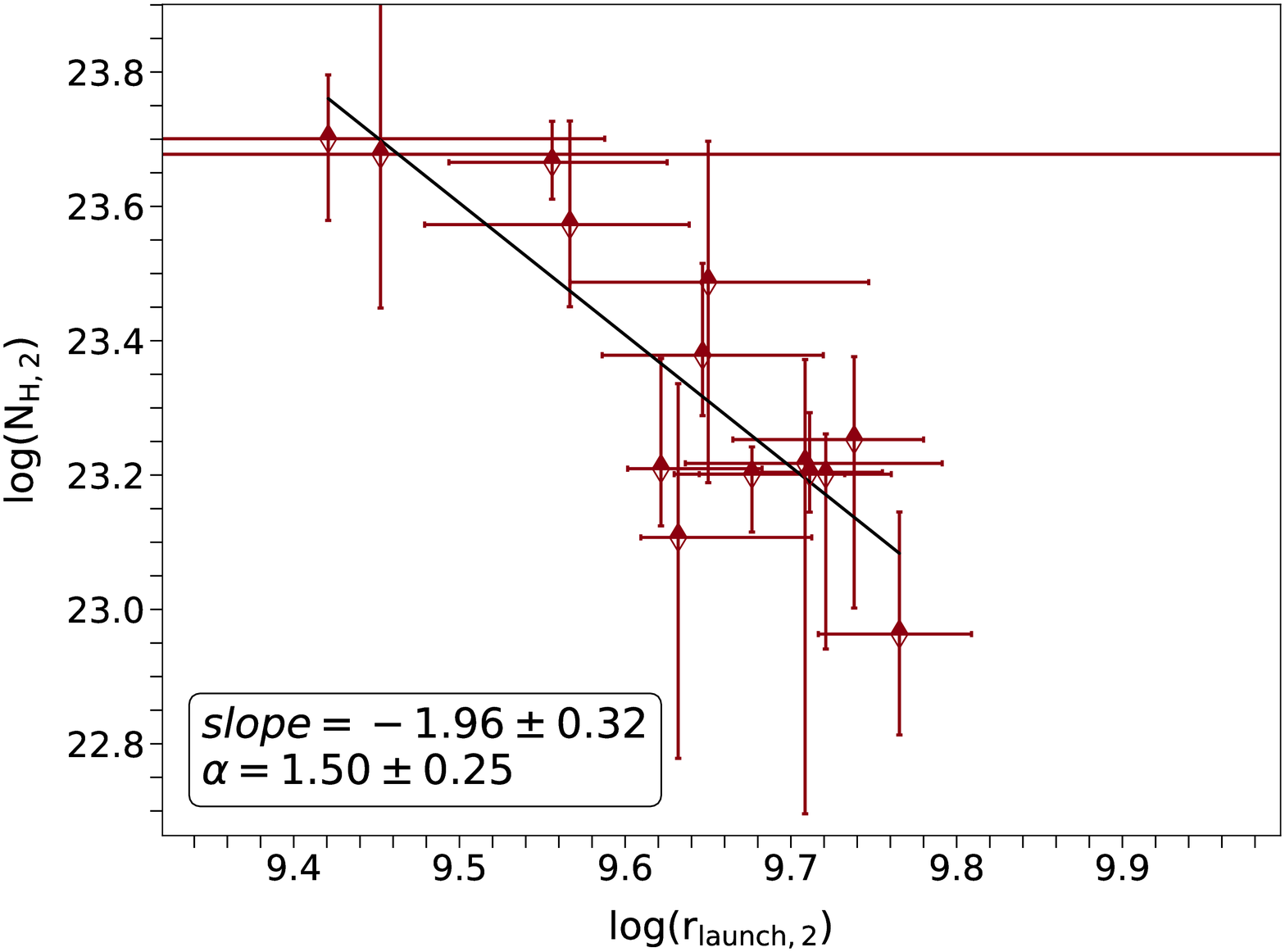}
\includegraphics[width=0.49\textwidth]{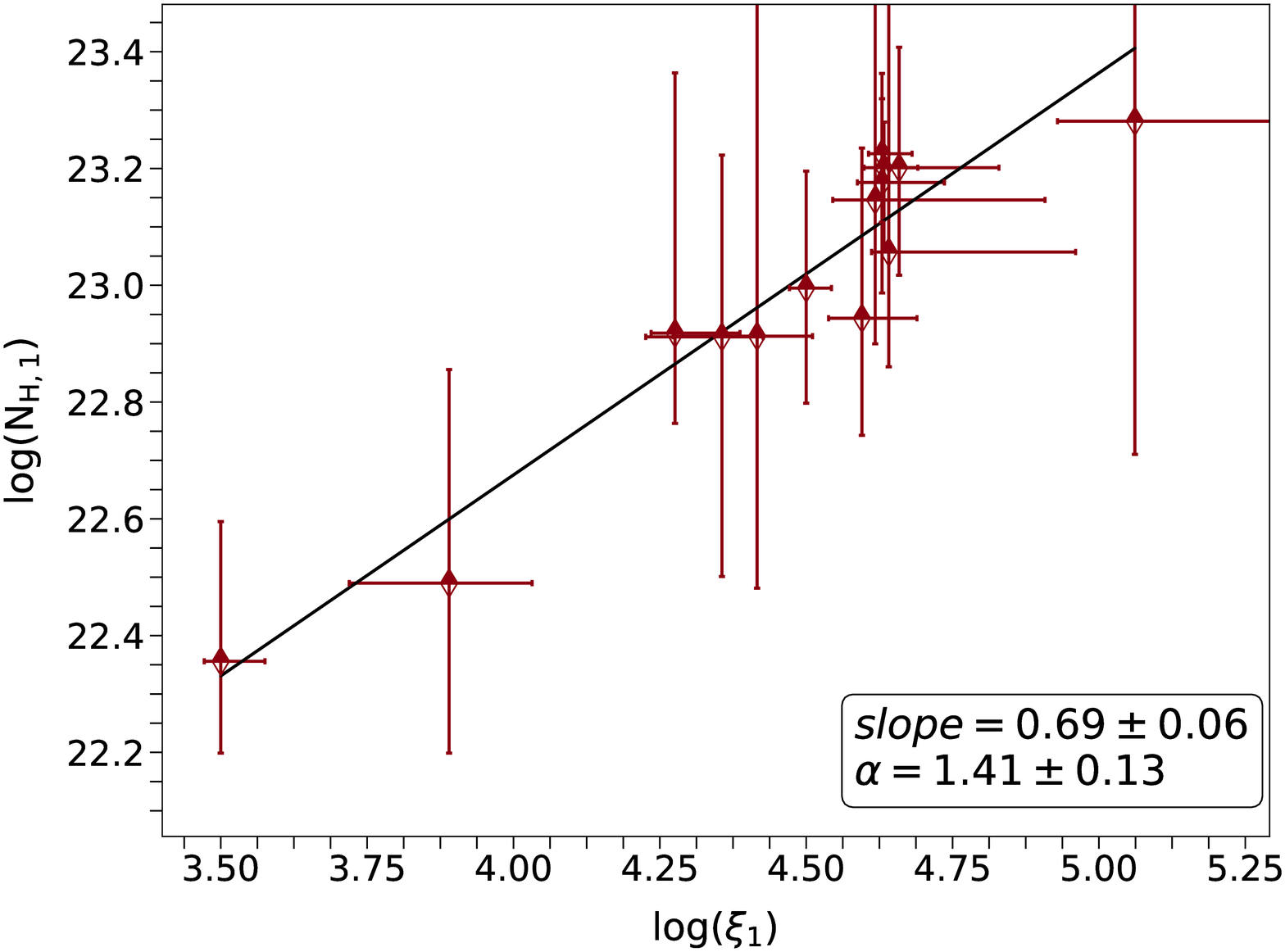}
\includegraphics[width=0.49\textwidth]{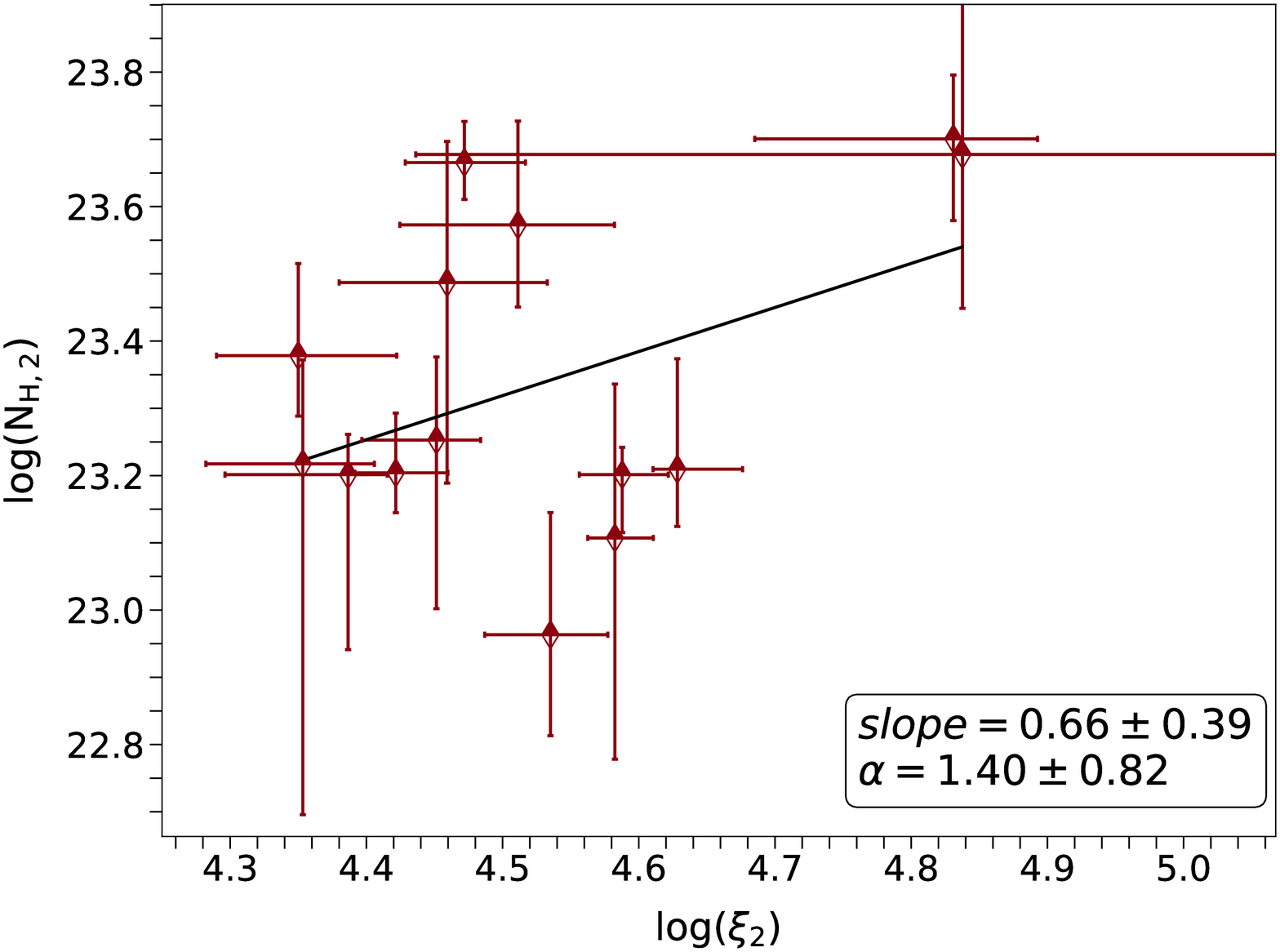}
\vspace{1mm}
\caption{The radial wind profile and Absorption Measure Distribution (AMD) for all absorbers (left: Zone 1; right: Zone 2).  The slopes from the radial wind profile (top two panels) give values of $\alpha = 1.37 \pm 0.19$ and $\alpha = 1.50 \pm 0.75$.  Fits to the AMD for each zone (bottom two panels) yield slopes that correspond to $\alpha = 1.41 \pm 0.13$ and $\alpha = 1.40 \pm 0.82$. The values of $\alpha$ contained within all of the error bounds are $1.28 \leq \alpha \leq 1.54)$.  These values are closest to the value of $\alpha = 1.4 $ from \citealt{2016A&A...589A.119C} but also encompass the value of $\alpha = 1.2$ \citep{2017NatAs...1E..62F} (except for the radial wind profile of the first absorber).  Broadly consistent values of $\alpha$ in Zone 1 and Zone 2 suggest a similar wind structure and consistent driving mechanism. }
\end{figure*}

\begin{figure*}[ht]
\centering
\includegraphics[width=0.49\textwidth]{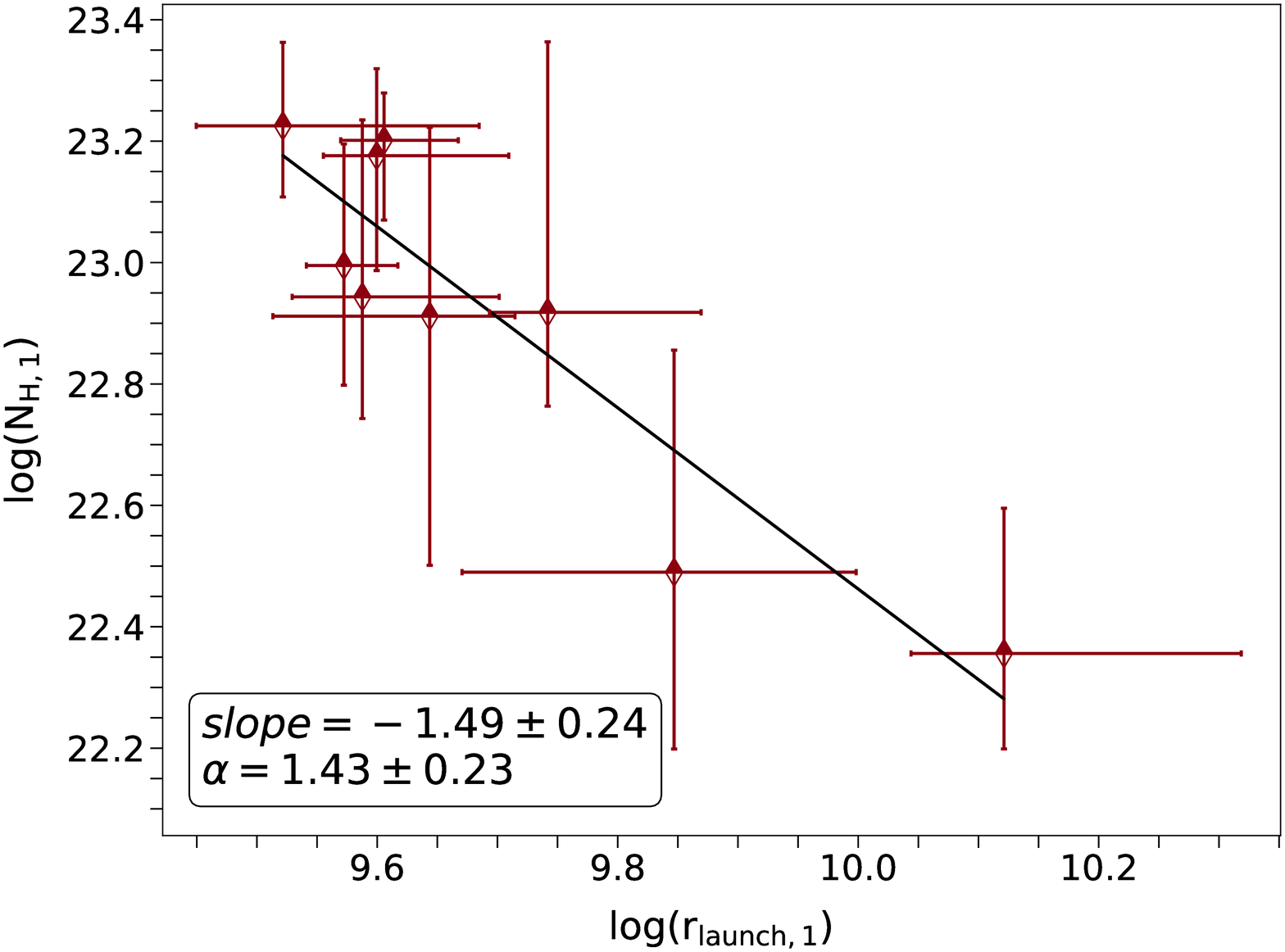}
\includegraphics[width=0.49\textwidth]{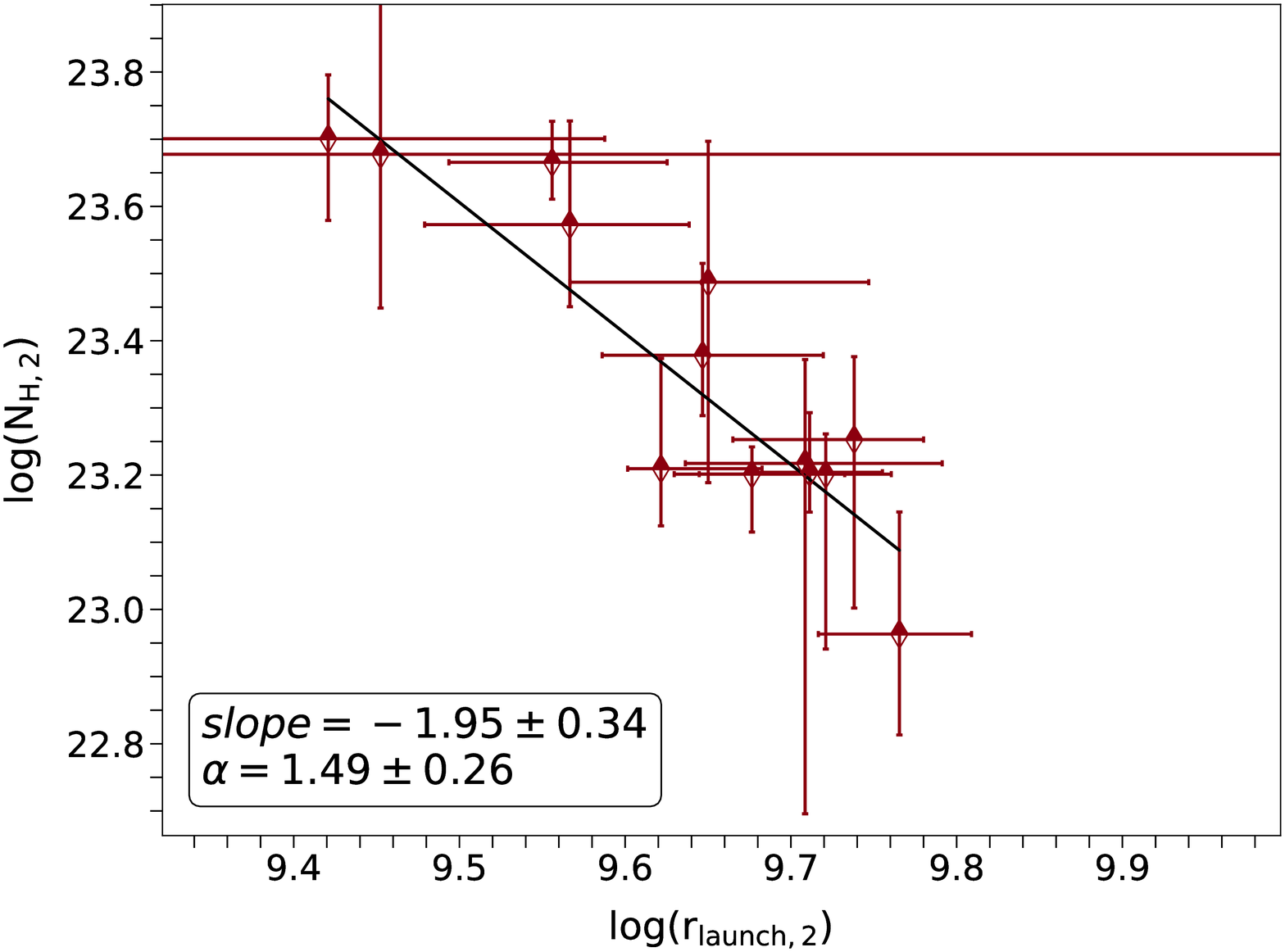}
\includegraphics[width=0.49\textwidth]{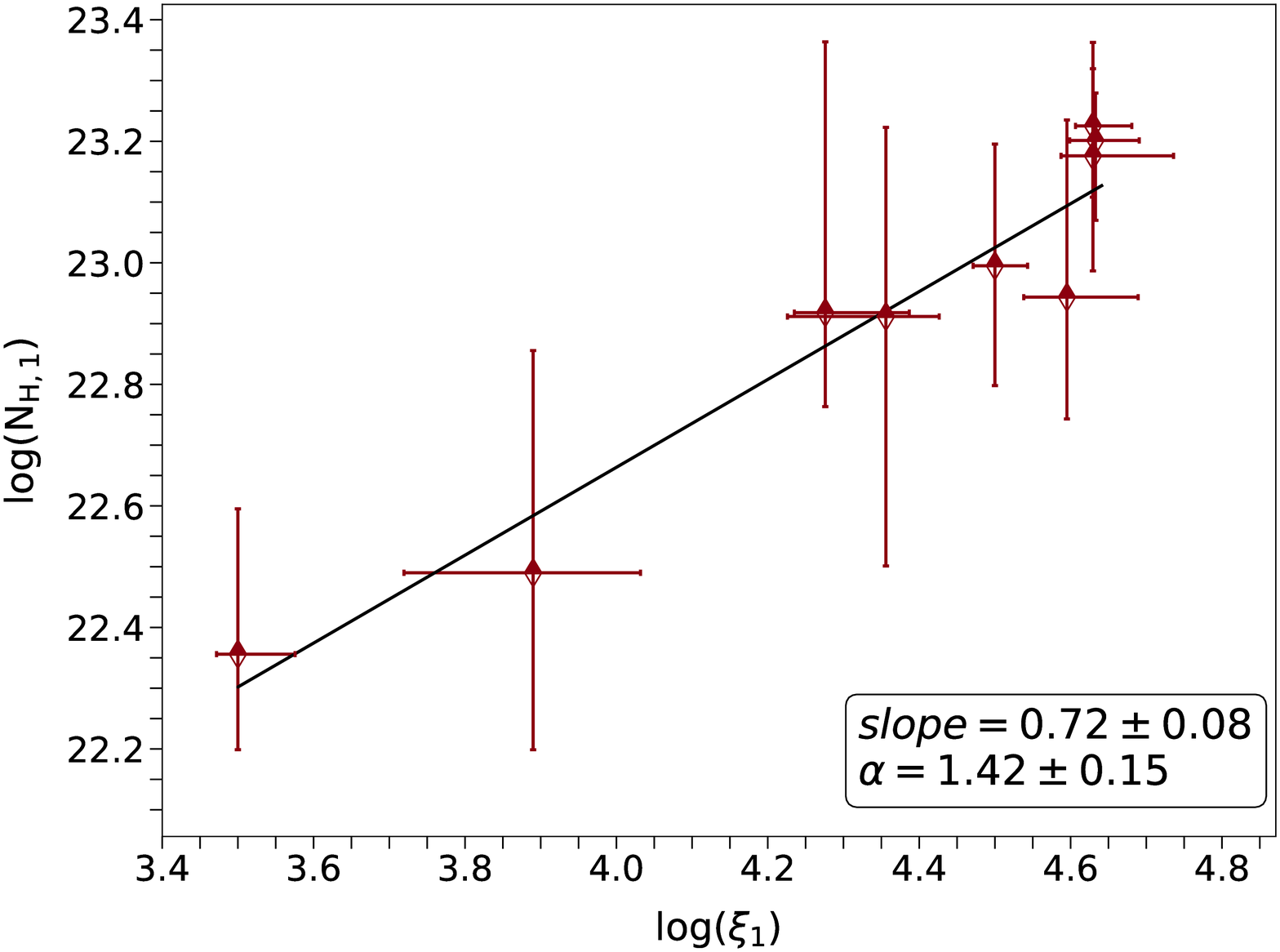}
\includegraphics[width=0.49\textwidth]{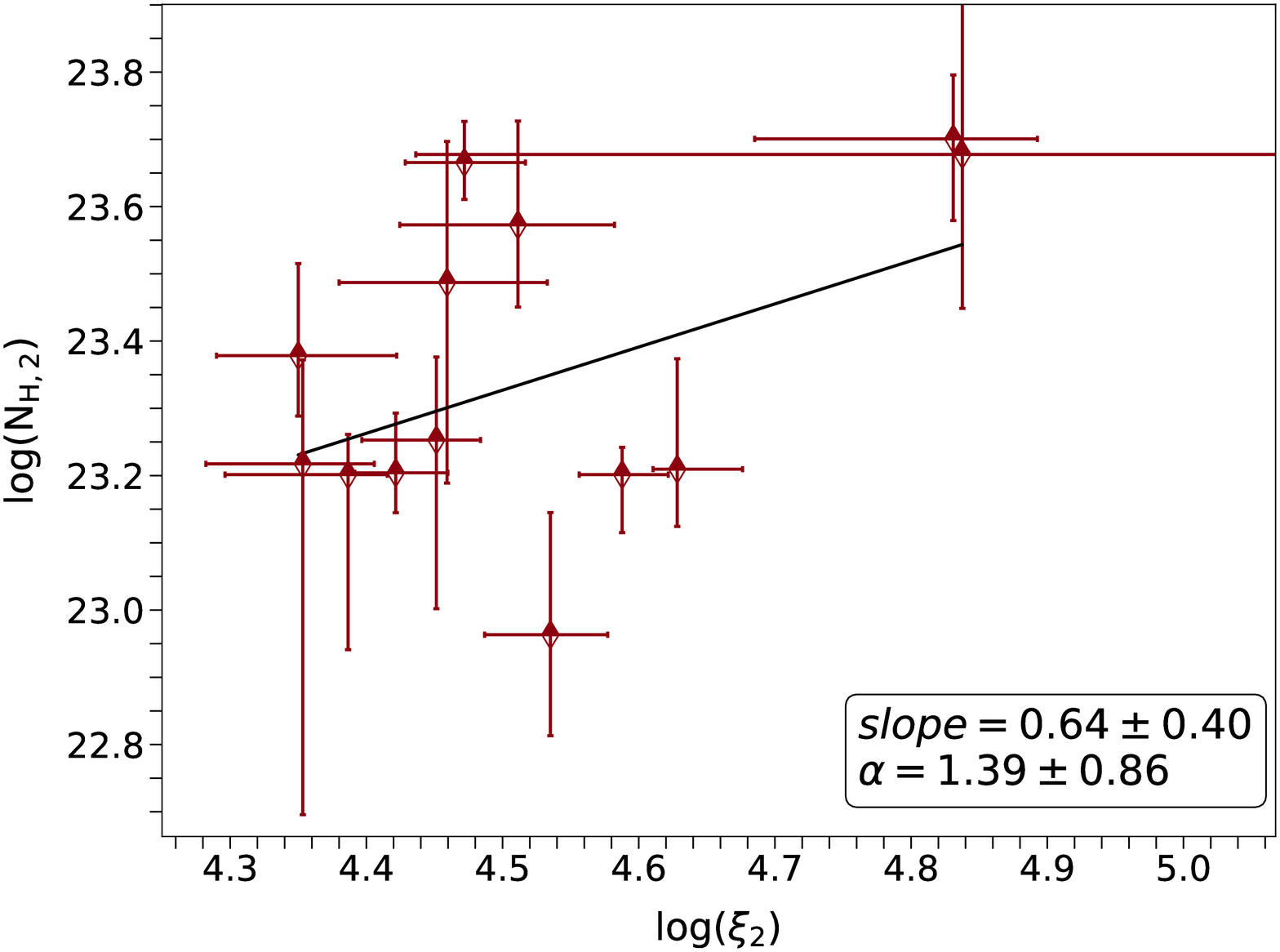}
\vspace{1mm}

\caption{Similar to Figure 8, but now excluding zones for which $N_{\rm H}-{\rm log}\xi$ error contours failed stringent quality metrics (see the main text).   The absolute values of $\alpha$ do not change significantly, but the uncertainty in $\alpha$ is increased by having fewer points.  The values of $\alpha$ contained within all the error bounds listed above are $1.27 \leq \alpha \leq 1.57$.}
\end{figure*} %

\begin{longrotatetable}
\begin{deluxetable*}{llllllllllll}
\tablewidth{700pt}
\tabletypesize{\scriptsize}
\tablecaption{Spectral Continuum Parameters}
\tablehead{
\colhead{ID} & \colhead{MJD} & 
\colhead{kT} & \colhead{Disk} & 
\colhead{$\Gamma$} & \colhead{PL} & 
\colhead{Observed} & \colhead{Unabsorbed} & 
\colhead{Luminosity} & \colhead{$\chi^2 / \nu$} & \colhead{$\chi^2 / \nu$} & \colhead{$\chi^2 / \nu$} \\ 
\colhead{} & \colhead{} & \colhead{(keV)} & \colhead{Norm} & 
\colhead{} & \colhead{Norm} & \colhead{Flux} &
\colhead{Flux} & \colhead{(10$^{37}$ erg/s)} & \colhead{No Wind} & \colhead{One Zone}  & \colhead{Two Zone}
} 
\startdata
00030009005 & 53448.0 & $1.269 \ \substack{+0.003 \\ -0.004} $ & $950 \ \substack{+10 \\ -10} $ & $1.00 \ \substack{+0.01 \\ -0.00} $ & $1.03 \ \substack{+0.02 \\ -0.01} $ & $4.7 \ \substack{+0.1  \\ -0.1} $ & $5.7 \ \substack{+0.1 \\ -0.1} $ & $5.7 \ \substack{+0.1 \\ -0.1} $ & 1426/736 = 1.938 & 1287/733 = 1.756 & 1072/730 = 1.468 \\
00030009006 & 53449.2 & $1.266 \ \substack{+0.003 \\ -0.005} $ & $930 \ \substack{+10 \\ -10} $ & $1.00 \ \substack{+0.01 \\ -0.00} $ & $1.02 \ \substack{+0.03 \\ -0.01} $ & $4.6 \ \substack{+0.1 \\ -0.1} $ & $5.5 \ \substack{+0.2 \\ -0.1} $ & $5.6 \ \substack{+0.2 \\ -0.1} $ & 1611/748 = 2.153 & 1433/745 = 1.923 & 1353/742 = 1.823 \\
00030009007 & 53450.2 & $1.294 \ \substack{+0.005 \\ -0.005} $ & $780 \ \substack{+10 \\ -10} $ & $1.00 \ \substack{+0.03 \\ -0.00} $ & $0.72 \ \substack{+0.05 \\ -0.02} $ & $4.0 \ \substack{+0.3 \\ -0.1} $ & $4.7 \ \substack{+0.3 \\ -0.1} $ & $4.9 \ \substack{+0.3 \\ -0.1} $ & 1335/710 = 1.880 & 1057/707 = 1.495 & 1053/704 = 1.496 \\
00030009008 & 53456.4 & $1.318 \ \substack{+0.006 \\ -0.006} $ & $420 \ \substack{+10 \\ -10} $ & $1.65 \ \substack{+0.03 \\ -0.02} $ & $1.68 \ \substack{+0.09 \\ -0.08} $ & $2.6 \ \substack{+0.1 \\ -0.1} $ & $3.1 \ \substack{+0.2 \\ -0.1} $ & $3.2 \ \substack{+0.2 \\ -0.1} $ & 1591/721 = 2.207 & 1331/718 = 1.854 & 1286/715 = 1.799 \\
00030009009 & 53463.5 & $1.300 \ \substack{+0.010 \\ -0.009} $ & $360 \ \substack{+10 \\ -10} $ & $2.30 \ \substack{+0.01 \\ -0.01} $ & $7.49 \ \substack{+0.07 \\ -0.06} $ & $2.9 \ \substack{+0.0 \\ -0.0} $ & $3.6 \ \substack{+0.0 \\ -0.0} $ & $3.5 \ \substack{+0.0 \\ -0.0} $ & 1256/689 = 1.823 & 1119/686 = 1.631 & 1170/683 = 1.713 \\
00030009010 & 53463.7 & $1.345 \ \substack{+0.008 \\ -0.008} $ & $410 \ \substack{+10 \\ -10} $ & $2.08 \ \substack{+0.01 \\ -0.01} $ & $4.68 \ \substack{+0.06 \\ -0.06} $ & $2.8 \ \substack{+0.0 \\ -0.0} $ & $3.6 \ \substack{+0.0 \\ -0.0} $ & $3.5 \ \substack{+0.0 \\ -0.0} $ & 1431/693 = 2.065 & 1149/690 = 1.853 & 1108/687 = 1.613 \\
00030009011 & 53470.4 & $1.238 \ \substack{+0.005 \\ -0.006} $ & $730 \ \substack{+10 \\ -20}$ & $1.31 \ \substack{+0.06 \\ -0.03} $ & $1.10 \ \substack{+0.17 \\ -0.07} $ & $3.3 \ \substack{+0.5 \\ -0.2} $ & $3.8 \ \substack{+0.6 \\ -0.3} $ & $4.0 \ \substack{+0.6 \\ -0.3} $ & 1307/692 = 1.889 & 1146/689 = 1.663 & 1091/686 = 1.590 \\
00030009012 & 53481.9 & $1.291 \ \substack{+0.020  \\ -0.020} $ & $750 \ \substack{+40 \\ -40} $ & $1.00 \ \substack{+0.07 \\ -0.00} $ & $0.65 \ \substack{+0.12 \\ -0.05} $ & $3.9 \ \substack{+0.7 \\ -0.3} $ & $4.5 \ \substack{+0.8 \\ -0.3} $ & $4.7 \ \substack{+0.9 \\ -0.3} $ & 539/456 = 1.182 & 506/453 = 1.117 & 487/450 = 1.082 \\
00030009014 & 53494.0 & $1.327 \ \substack{+0.004 \\ -0.004} $ & $710 \ \substack{+10 \\ -10} $ & $1.00 \ \substack{+0.03 \\ -0.00} $ & $0.66 \ \substack{+0.04 \\ -0.01} $ & $4.1 \ \substack{+0.3 \\ -0.1} $ & $4.8 \ \substack{+0.3 \\ -0.1} $ & $5.0 \ \substack{+0.3 \\ -0.1} $ & 1155/682 = 1.694 & 1000/679 = 1.473 & 922/676 = 1.363 \\
00030009015 & 53504.3 & $1.201 \ \substack{+0.005 \\ -0.005} $ & $1200 \ \substack{+20 \\ -20} $ & $1.20 \ \substack{+0.01 \\ -0.02} $ & $2.56 \ \substack{+0.07 \\ -0.12} $ & $5.6 \ \substack{+0.2 \\ -0.3} $ & $6.9 \ \substack{+0.2 \\ -0.3} $ & $6.8 \ \substack{+0.2 \\ -0.3} $ & 1971/760 = 2.593 & 1788/757 = 2.362 & 1699/754 = 2.253 \\
00030009016 & 53505.4 & $1.279 \ \substack{+0.004 \\ -0.005} $ & $940 \ \substack{+10 \\ -10} $ & $1.00 \ \substack{+0.01 \\ -0.00} $ & $1.05 \ \substack{+0.04 \\ -0.01} $ & $5.2 \ \substack{+0.2 \\ -0.1} $ & $6.1 \ \substack{+0.2 \\ -0.1} $ & $6.4 \ \substack{+0.2 \\ -0.1} $ & 1000/704 = 2.593 & 886/701 = 1.263 & 863/697 = 1.238 \\
00030009017 & 53506.5 & $1.171 \ \substack{+0.004 \\ -0.004}$ & $1700 \ \substack{+20 \\ -20} $ & $1.00 \ \substack{+0.01 \\ -0.00} $ & $3.18 \ \substack{+0.04 \\ -0.02} $ & $8.0 \ \substack{+0.1 \\ -0.1} $ & $9.5 \ \substack{+0.1 \\ -0.1} $ & $9.7 \ \substack{+0.1 \\ -0.1} $ & 1400/791 = 1.770 & 1357/788 = 1.722 & 1276/785 = 1.625 \\
00030009018 & 53511.4 & $1.239 \ \substack{+0.004 \\ -0.003}$ & $1200 \ \substack{+10 \\ -10} $ & $1.00 \ \substack{+0.01 \\ -0.00} $ & $1.80 \ \substack{+0.08 \\ -0.01} $ & $6.0 \ \substack{+0.3 \\ -0.0} $ & $7.1 \ \substack{+0.3 \\ -0.0} $ & $7.4 \ \substack{+0.3 \\ -0.0} $ & 2237/790 = 2.832 & 2126/787 = 2.701 & 1967/784 = 2.509 \\
00030009019 & 53512.3 & $1.311 \ \substack{+0.005 \\ -0.003}$ & $850 \ \substack{+0 \\ -10} $ & $1.00 \ \substack{+0.03 \\ -0.00} $ & $1.03 \ \substack{+0.08 \\ -0.01} $ & $4.8 \ \substack{+0.4 \\ -0.1} $ & $5.7 \ \substack{+0.41 \\ -0.1} $ & $5.9 \ \substack{+0.5 \\ -0.1} $ & 1734/762 = 2.276 & 1523/759 = 2.007 & 1131/754 = 1.500 \\
00030009020 & 53512.9 & $1.319 \ \substack{+0.005 \\ -0.007} $ & $880 \ \substack{+10 \\ -20} $ & $1.00 \ \substack{+0.02 \\ -0.00} $ & $0.98 \ \substack{+0.07 \\ -0.02} $ & $5.0 \ \substack{+0.3 \\ -0.1} $ & $5.9 \ \substack{+0.4 \\ -0.1} $ & $6.1 \ \substack{+0.4 \\ -0.1} $ & 928/638 = 1.454 & 856/635 = 1.348 & 821/632 = 1.299
\enddata
\tablecomments{The table above lists values from the continuum fits, including the temperature, disk normalization, photon index, and powerlaw normalization.  The photon index was constrained to physical values, $1 \leq \Gamma \leq 4$. We list the 1-9 keV flux. The parameter values are from observations treated with the two-absorber model. The unabsorbed flux was calculated by freezing the internal obscuration at 0 and calculating the flux. The $\chi^2/\nu$ values are listed for each observation with no absorber, one absorber, or two absorbers. In all observations, adding an absorber component reduced the $\chi^2$ value. }
\end{deluxetable*}
\end{longrotatetable}

\begin{longrotatetable}
\begin{deluxetable*}{llrrrrrrll}
\tablewidth{700pt}
\tabletypesize{\scriptsize}
\tablecaption{Measured Wind Absorption Parameters}
\tablehead{
\colhead{ID} & \colhead{MJD} & 
\colhead{Column} & \colhead{$\log(\xi)_1$} & \colhead{Velocity I} & 
\colhead{Column} & \colhead{$\log(\xi)_2$} & \colhead{Velocity II} \\ 
\colhead{} & \colhead{} & \colhead{Density I} & \colhead{} & 
\colhead{} & \colhead{Density II} &
\colhead{} & \colhead{}
\\
\colhead{} & \colhead{} & \colhead{(10$^{22}$ cm$^{-2}$)} & \colhead{} & 
\colhead{($10^3$ km/s)} & \colhead{(10$^{22}$ cm$^{-2}$)} &
\colhead{} & \colhead{(km/s)} 
}
\startdata
00030009005 & 53448.0 & $16 \ \substack{+1 \\ -2} $ & $4.63 \ \substack{+0.06 \\ -0.03} $ & $-10 \ \substack{+1 \\ -1} $ & $16 \ \substack{+1 \\ -1} $ & $4.42 \ \substack{+0.04 \\ -0.03} $ & $-900 \ \substack{+400 \\ -400} $ \\
00030009006 & 53449.2 & $15 \ \substack{+2 \\ -3} $ & $4.63 \ \substack{+0.11 \\ -0.04} $ & $-9 \ \substack{+1 \\ -1} $ & $16 \ \substack{+1 \\ -4} $ & $4.39 \ \substack{+0.03 \\ -0.09} $ & $-800 \ \substack{+400 \\ -300} $ \\
00030009007 & 53450.2 & $14 \ \substack{+7 \\ -4} $ & $4.62 \ \substack{+0.29 \\ -0.07} $ & $-6 \ \substack{+1 \\ -1} $ & $31 \ \substack{+6 \\ -9} $ & $4.47 \ \substack{+0.07 \\ -0.06} $ & $-600 \ \substack{+500  \\ -500} $ \\
00030009008 & 53456.4 & $8 \ \substack{+4 \\ -1} $ & $4.28 \ \substack{+0.11 \\ -0.04} $ & $-23 \ \substack{+1 \\ -0} $ & $46 \ \substack{+3 \\ -3} $ & $4.47 \ \substack{+0.04 \\ -0.04} $ & $-400 \ \substack{+300 \\ -400} $ \\
00030009009 & 53463.5 & $8 \ \substack{+5 \\ -4} $ & $4.42 \ \substack{+0.11 \\ -0.04} $ & $-20 \ \substack{+1 \\ -1} $ & $24 \ \substack{+3 \\ -2} $ & $4.35 \ \substack{+0.07 \\ -0.06} $ & $-700 \ \substack{+700 \\ -500} $ \\
00030009010 & 53463.7 & $3 \ \substack{+1 \\ -1} $ & $3.89 \ \substack{+0.11 \\ -0.04} $ & $-30 \ \substack{+1 \\ -2} $ & $37 \ \substack{+6 \\ -5} $ & $4.51 \ \substack{+0.07 \\ -0.09} $ & $-2200 \ \substack{+400 \\ -380} $ \\
00030009011 & 53470.4 & $8 \ \substack{+3 \\ -3} $ & $4.36 \ \substack{+0.07 \\ -0.13} $ & $-20 \ \substack{+1 \\ -1} $ & $50 \ \substack{+5 \\ -6} $ & $4.83 \ \substack{+0.06 \\ -0.15 } $ & $-1300 \ \substack{+400 \\ -400} $ \\
00030009012 & 53481.9 & $19 \ \substack{+12 \\ -11} $ & $5.06 \ \substack{+0.57 \\ -0.13} $ & $-40 \ \substack{+1 \\ -1} $ & $50 \ \substack{+12 \\ -11} $ & $4.84 \ \substack{+0.52 \\ -0.40} $ & $-1400 \ \substack{+800  \\ -1000} $ \\
00030009014 & 53494.0 & $16 \ \substack{+3 \\ -3} $ & $4.66 \ \substack{+0.17 \\ -0.02} $ & $-9 \ \substack{+2 \\ -1} $ & $17 \ \substack{+3 \\ -9} $ & $4.35 \ \substack{+0.05 \\ -0.07} $ & $-110 \ \substack{+80 \\ -1000} $ \\
00030009015 & 53504.3 & $10 \ \substack{+2 \\ -2} $ & $4.50 \ \substack{+0.04 \\ -0.03} $ & $-9 \ \substack{+1 \\ -1} $ & $18 \ \substack{+2 \\ -5} $ & $4.45 \ \substack{+0.03 \\ -0.05} $ & $-100 \ \substack{+70 \\ -1100} $ \\
00030009016 & 53505.4 & $17 \ \substack{+2 \\ -2} $ & $4.63 \ \substack{+0.05 \\ -0.02} $ & $-5 \ \substack{+1 \\ -1} $ & $16 \substack{+3 \\ -1} $ & $4.63 \ \substack{+0.05 \\ -0.02} $ & $-100 \ \substack{+70 \\ -700} $ \\
00030009017 & 53506.5 & $3 \ \substack{+1 \\ -1} $ & $3.50 \ \substack{+0.08 \\ -0.03} $ & $-25\ \substack{+1 \\ -1} $ & $9 \ \substack{+2 \\ -1} $ & $4.54 \ \substack{+0.04 \\ -0.05} $ & $-590 \ \substack{+500 \\ -900} $ \\
00030009018 & 53511.4 & $9 \ \substack{+3\\ -2} $ & $4.60 \ \substack{+0.09 \\ -0.06} $ & $-7 \ \substack{+1 \\ -1} $ & $16 \ \substack{+1 \\ -1} $ & $4.59 \ \substack{+0.03 \\ -0.03} $ & $-100 \ \substack{+100 \\ -1000} $ \\
00030009019 & 53512.3 & $11 \ \substack{+5 \\ -2} $ & $4.64 \ \substack{+0.32 \\ -0.03} $ & $-7 \ \substack{+1 \\ -1} $ & $13 \ \substack{+3 \\ -4} $ & $4.58 \ \substack{+0.03 \\ -0.02} $ & $-500 \ \substack{+500 \\ -900} $ \\
00030009020 & 53512.9 & $13 \ \substack{+5 \\ -7} $ & $4.58 \ \substack{+0.16 \\ -0.18} $ & $-11 \ \substack{+1 \\ -1} $ & $16 \ \substack{+4 \\ -2} $ & $4.49 \ \substack{+0.07 \\ -0.03} $ & $-800 \ \substack{+600 \\ -600} $
\enddata
\tablecomments{The table above lists the best-fit parameters for the properties of the wind components. The column density fluctuates throughout the outburst. The ionisation is well constrained in most cases, and is generally high ($\rm \log{\xi} > 4.27$), other than two fits which give $\rm \log{\xi} = 3.89$ and $3.5$.  We note that OBSID 0003009012 has higher errors than the other observations}. The velocities have been corrected for gravitational red-shift). It should be noted that the highest velocities are very well constrained, while lower velocities are not always well constrained given the limited resolution of the CCD. The errors listed for all values are 1 $\sigma$ errors.
\end{deluxetable*}
\end{longrotatetable}

\begin{longrotatetable}
\begin{deluxetable*}{llrrrrrrrll}
\tablewidth{700pt}
\tabletypesize{\scriptsize}
\tablecaption{Radii}
\tablehead{
\colhead{ID} & \colhead{MJD} & \colhead{$R_C$} & \colhead{$\rm \log{(r_{photo, 1})}$}  & \colhead{$\rm \log{(r_{launch, 1})}$} & \colhead{$\rm \log{(r_{photo, 2})}$}  & \colhead{$\rm \log{(r_{launch, 2})}$}  \\ 
}
\startdata
00030009005 & 53448.0 & $11.677 \ \substack{+0.003 \\ -0.003} $ & $10.0 \ \substack{+0.1 \\ -0.1} $ & $9.61 \ \substack{+0.06 \\ -0.04} $ & $10.2 \ \substack{+0.1 \\ -0.1} $ & $9.71 \ \substack{+0.04 \\ -0.03} $ \\
00030009006 & 53449.2 & $11.678 \ \substack{+0.002 \\ -0.004} $ & $10.0 \ \substack{+0.2 \\ -0.2} $ & $9.60 \ \substack{+0.11 \\ -0.04} $ & $10.2 \ \substack{+0.1 \\ -0.3} $ & $9.72 \ \substack{+0.04 \\ -0.09} $ \\
00030009007 & 53450.2 & $11.668 \ \substack{+0.004 \\ -0.004} $ & $10.0 \ \substack{+0.6 \\ -0.3} $ & $9.57 \ \substack{+0.30 \\ -0.08} $ & $9.8 \ \substack{+0.2 \\ -0.3} $ & $9.65 \ \substack{+0.10 \\ -0.08} $ \\
00030009008 & 53456.4 & $11.660 \ \substack{+0.004 \\ -0.004} $ & $10.4 \ \substack{+0.5 \\ -0.2} $ & $9.65 \ \substack{+0.12 \\ -0.06} $ & $9.5 \ \substack{+0.1 \\ -0.1} $ & $9.56 \ \substack{+0.07 \\ -0.06} $ \\
00030009009 & 53463.5 & $11.667 \ \substack{+0.008 \\ -0.007} $ & $10.3 \ \substack{+0.6 \\ -0.5} $ & $9.61 \ \substack{+0.10 \\ -0.14} $ & $9.9 \ \substack{+0.2 \\ -0.1} $ & $9.65 \ \substack{+0.07 \\ -0.06} $ \\
00030009010 & 53463.7 & $11.652 \ \substack{+0.006 \\ -0.006} $ & $11.3 \ \substack{+0.4 \\ -0.4} $ & $9.88 \ \substack{+0.14 \\ -0.17} $ & $9.6 \ \substack{+0.2 \\ -0.2} $ & $9.57 \ \substack{+0.07 \\ -0.09} $ \\
00030009011 & 53470.4 & $11.688 \ \substack{+0.004 \\ -0.005} $ & $10.4 \ \substack{+0.4 \\ -0.4} $ & $9.66 \ \substack{+0.17 \\ -0.17} $ & $9.1 \ \substack{+0.2 \\ -0.2} $ & $9.42 \ \substack{+0.17 \\ -0.16} $ \\
00030009012 & 53481.9 & $11.670 \ \substack{+0.012 \\ -0.012} $ & $9.4 \ \substack{+0.9 \\ -0.6} $ & $9.34 \ \substack{+0.60 \\ -0.15} $ & $9.2 \ \substack{+0.6 \\ -0.5} $ & $9.45 \ \substack{+0.55 \\ -0.41} $ \\
00030009014 & 53494.0 & $11.657\ \substack{+0.003 \\ -0.003} $ & $9.9 \ \substack{+0.3 \\ -0.2} $ & $9.56 \ \substack{+0.18 \\ -0.02} $ & $10.2 \ \substack{+0.2 \\ -0.5} $ & $9.71 \ \substack{+0.08 \\ -0.07} $ \\
00030009015 & 53504.3 & $11.701 \ \substack{+0.004 \\ -0.004} $ & $10.4 \ \substack{+0.2 \\ -0.2} $ & $9.71 \ \substack{+0.05 \\ -0.06} $ & $10.2 \ \substack{+0.1 \\ -0.3} $ & $9.74 \ \substack{+0.04 \\ -0.07} $ \\
00030009016 & 53505.4 & $11.674 \ \substack{+0.004 \\ -0.004},$ & $10.0 \ \substack{+0.2 \\ -0.1} $ & $9.62 \ \substack{+0.06 \\ -0.02} $ & $10.0 \ \substack{+0.2 \\ -0.1} $ & $9.62 \ \substack{+0.06 \\ -0.02} $ \\
00030009017 & 53506.5 & $11.712 \ \substack{+0.003 \\ -0.003} $ & $12.2 \ \substack{+0.3 \\ -0.2} $ & $10.28 \ \substack{+0.08 \\ -0.03} $ & $10.6 \ \substack{+0.2 \\ -0.2} $ & $9.77 \ \substack{+0.04 \\ -0.05} $ \\
00030009018 & 53511.4 & $11.687 \ \substack{+0.003 \\ -0.002} $ & $10.4 \ \substack{+0.3 \\ -0.2} $ & $9.67 \ \substack{+0.10 \\ -0.06} $ & $10.2 \ \substack{+0.1 \\ -0.1} $ & $9.68 \ \substack{+0.06 \\ -0.03} $ \\
00030009019 & 53512.3 & $11.663 \ \substack{+0.003 \\ -0.002} $ & $10.2 \ \substack{+0.6 \\ -0.2} $ & $9.60 \ \substack{+0.33 \\ -0.03} $ & $10.2 \ \substack{+0.2 \\ -0.3} $ & $9.63 \ \substack{+0.08 \\ -0.02} $ \\
00030009020 & 53512.9 & $11.66 \ \substack{+0.004 \\ -0.005} $ & $10.2 \ \substack{+0.4 \\ -0.5} $ & $9.64 \ \substack{+0.17 \\ -0.18} $ & $10.2 \ \substack{+0.3 \\ -0.1} $ & $9.69 \ \substack{+0.01 \\ -0.04} $
\enddata
\tablecomments{Important radii calculated in different ways.  We list values of the Compton radius, photoionisation radius, and launching radius for both zones.  The statistical errors on the Compton radius are small due to the fact that uncertainties on the disk temperature are small; systematic errors are likely to be larger.  Recall the photoionisation radius is a strict upper limit, calculated using the column density of the absorber, while the launching radius was calculated using $n \approx 10^{14} \rm cm^{-3}$.  The radius is in log units of cm.  The derived radii are consistently at least an order of magnitude below the Compton radius, implying that the observed outflows cannot be driven by a thermal wind.}
\end{deluxetable*}
\end{longrotatetable}

\begin{longrotatetable}
\begin{deluxetable*}{llrrrrrrrll}
\tablewidth{700pt}
\tabletypesize{\scriptsize}
\tablecaption{Derived Wind Properties}
\tablehead{
\colhead{ID} & \colhead{MJD} & \colhead{$\dot{M}_{wind}$ I} & \colhead{$L_{kin}$ I}  & \colhead{$\dot{M}_{wind}$ II} & \colhead{$L_{kin}$ II} &  \\ 
\colhead{}  & \colhead{} & \colhead{(10$^{18}$ g s$^{-1}$)} & \colhead{(10$^{36}$ ergs)}  & \colhead{(10$^{18}$ g s$^{-1}$)}  & \colhead{(10$^{31}$ ergs)}
}
\startdata
00030009005 & 53448.0 & $4.3 \ \substack{+0.4 \\ -0.4} $ & $1.8  \ \substack{+0.3 \\ -0.3} $ & $0.6 \ \substack{+0.3 \\ -0.3} $ & $210 \substack{+180 \\ -160} $ \\
00030009006 & 53449.2 & $3.9 \ \substack{+0.7 \\ -0.3} $ & $1.5 \ \substack{+0.4 \\ -0.2} $ & $0.7 \ \substack{+0.3 \\ -0.3} $ & $220 \ \substack{+170 \\ -140} $ \\
00030009007 & 53450.2 & $2.1 \ \substack{+0.7 \\ -0.6} $ & $0.3 \ \substack{+0.1 \\ -0.1} $ & $0.3\ \substack{+0.3 \\ -0.3} $ & $50 \ \substack{+80 \\ -70} $ \\
00030009008 & 53456.4 & $14 \ \substack{+2 \\ -1} $ & $37  \ \substack{+5 \\ -3} $ & $0.1 \ \substack{+0.1 \\ -0.1} $ & $7 \ \substack{+10 \\ -7} $ \\
00030009009 & 53463.5 & $10 \ \substack{+1 \\ -2} $ & $20 \ \substack{+3 \\ -4} $ & $0.4 \ \substack{+0.4 \\ -0.3} $ & $90 \ \substack{+160 \\ -90} $ \\
00030009010 & 53463.7 & $48 \ \substack{+7 \\ -9} $ & $210 \ \substack{+30 \\ -40} $ & $0.9  \ \substack{+0.2 \\ -0.2} $ & $2200 \ \substack{+700 \\ -700} $ \\
00030009011 & 53470.4 & $14 \ \substack{+3 \\ -2} $ & $40 \ \substack{+8 \\ -7} $ & $0.3 \ \substack{+0.1 \\ -0.1} $ & $200 \ \substack{+100 \\ -100} $ \\
00030009012 & 53481.9 & $6.1 \ \substack{+3.7 \\ -0.9} $ & $70 \ \substack{+40 \\ -10} $ & $0.3 \ \substack{+0.2 \\ -0.3} $ & $280 \ \substack{+310 \\ -280} $ \\
00030009014 & 53494.0 & $3.1  \ \substack{+0.8 \\ -0.3} $ & $1.1 \ \substack{+0.4 \\ -0.2} $ & $0.1 \ \substack{+0.1 \\ -0.1} $ & $0.2 \ \substack{+0.4 \\ -0.2} $ \\
00030009015 & 53504.3 & $6.5 \ \substack{+0.4 \\ -1.0} $ & $2.4 \ \substack{+0.2 \\ -0.6} $ & $0.1 \ \substack{+0.1 \\ -0.1} $ & $0.2 \ \substack{+0.2 \\ -0.2} $ \\
00030009016 & 53505.4 & $2.7 \ \substack{+0.3 \\ -0.2} $ & $0.4 \ \substack{+0.1 \\ -0.1} $ & $0.1 \ \substack{+0.1 \\ -0.1} $ & $0.1 \ \substack{+0.1 \\ -0.1} $ \\
00030009017 & 53506.5 & $260 \ \substack{+20 \\ -10} $ & $780 \ \substack{+70 \\ -60} $ & $0.6 \ \substack{+0.5 \\ 0.6} $ & $90 \ \substack{+130 \\ -90} $ \\
00030009018 & 53511.4 & $4.3 \ \substack{+0.8 \\ -0.5} $ & $1.0 \ \substack{+0.3 \\ -0.2} $ & $0.1 \ \substack{+0.1 \\ -0.1} $ & $0.1 \ \substack{+0.2 \\ -0.1} $ \\
00030009019 & 53512.3 & $3.0 \ \substack{+1.1 \\ -0.6} $ & $0.6 \ \substack{+0.3 \\ -0.2} $ & $0.3 \ \substack{+0.3 \\ -0.3} $ & $30 \ \substack{+50 \\ -30} $ \\
00030009020 & 53512.9 & $5.8 \ \substack{+1.1 \\ -1.1} $ & $3.4 \ \substack{+0.7 \\ -0.7} $ & $0.6 \ \substack{+0.4 \\ -0.4} $ & $180 \ \substack{+220 \\ -180} $
\enddata
\tablecomments{Derived quantities from our best-fit parameters. Note that the kinetic power for Zone 2 is in units of $10^{31} \rm ergs$ while Zone 1 is in units of $10^{36} \rm ergs$ to maintain consistency with precision (since $L_{kin} \propto v^{3}$). However, the values in this table should effectively serve as upper limits, as we did not calculate a single filling factor to use in our calculations of mass outflow rate.  Some errors on the mass outflow rate and kinetic power in Zone 2 are large owing to similar uncertainties in the outflow velocity for Zone 2. There are several extremal values of kinetic power and mass outflow rate that are significantly lower than the other values; these are due to the variations in the measured velocities.}
\end{deluxetable*}
\end{longrotatetable}

\appendix

\section{Formulations of Relevant Correlations}

The purpose of this appendix is to show the work that led us to our formulations for the relationships between column density and launching radius, and column density and ionization parameter.

Using $n_{14} = 10^{14} cm^{-3}$ and the definition of the column density, $N_H$:
\begin{equation}
N_H = n_{14} \Delta r_{14} = n_{14} \frac{\Delta r_{14}}{r_{14}} r_{14} = n_{14} f_{14} r_{14}
\end{equation}

And the equation for launching radius:
\begin{equation}
r^2 = \frac{L}{\xi n} = \frac{L}{\xi n_{14}} \frac{n_{14}}{n} = r_{14}^2 \frac{n_{14}}{n}
\end{equation}

Let us define the density as $n = n_0 r^{-\alpha}$. Then (A1) and (A2) become:

\begin{equation}
N_H = f n_0 r^{1-\alpha} 
\end{equation}
\begin{equation}
r^2 = r_{14}^2 \frac{n_{14}}{n_0 r^{-\alpha}} \longrightarrow r = \Big( r_{14}^2 \frac{n_{14}}{n_0} \Big)^{1/(2-\alpha)}
\end{equation}

Solving for r in (A4) and plugging it into (A3) gives us:

\[ N_H = \Big(\frac{n_0}{n_{14}} \Big)^{\frac{1-\alpha}{2-\alpha}} n_0 f r_{14}^{\frac{2-2\alpha}{2-\alpha}} \]

Similarly, we can start from (A4) to derive a similar expression involving $\xi$ by substituting in the definition of launching radius.

\begin{equation}
r = \Big( r_{14}^2 \frac{n_{14}}{n_0} \Big)^{1/(2-\alpha)} =  \Big( \frac{L}{\xi_{14} n_{14}} \frac{n_{14}}{n_0} \Big)^{1/(2-\alpha)} = \xi_{14}^{-\frac{1}{2-\alpha}} (L_{14} n_0)^{\frac{1}{2-\alpha}}
\end{equation}

Plugging (A5) into (A3) gives us:

\[ N_H = (L_{14} n_0)^{\frac{1-\alpha}{2-\alpha}} n_0 \ f \ \xi_{14}^{\frac{\alpha-1}{2-\alpha}} \]

Therefore, we used $\rm exponent = \frac{2-2\alpha}{2-\alpha}$ to calculate alpha from the measured coefficient in the radial wind profiles. We then used $\rm exponent = \frac{\alpha-1}{2-\alpha} $ to calculate alpha from the measured coefficient in the Absorption Measure Distributions.

\bibliography{ref}{}

\begin{thebibliography}{}
\expandafter\ifx\csname natexlab\endcsname\relax\def\natexlab#1{#1}\fi
\providecommand{\url}[1]{\href{#1}{#1}}

\bibitem[{{Arnaud}(1996)}]{1996ASPC..101...17A}
{Arnaud}, K.~A. 1996, in Astronomical Society of the Pacific Conference Series,
  Vol. 101, Astronomical Data Analysis Software and Systems V, ed. G.~H.
  {Jacoby} \& J.~{Barnes}, 17

\bibitem[{{Balbus} \& {Hawley}(1991)}]{1991ApJ...376..214B}
{Balbus}, S.~A., \& {Hawley}, J.~F. 1991, \apj, 376, 214

\bibitem[{{Begelman} {et~al.}(2015){Begelman}, {Armitage}, \&
  {Reynolds}}]{2015ApJ...809..118B}
{Begelman}, M.~C., {Armitage}, P.~J., \& {Reynolds}, C.~S. 2015, \apj, 809, 118

\bibitem[{{Begelman} {et~al.}(1983){Begelman}, {McKee}, \&
  {Shields}}]{1983ApJ...271...70B}
{Begelman}, M.~C., {McKee}, C.~F., \& {Shields}, G.~A. 1983, \apj, 271, 70

\bibitem[{{Behar}(2009)}]{2009ApJ...703.1346B}
{Behar}, E. 2009, \apj, 703, 1346

\bibitem[{{Brocksopp} {et~al.}(2006){Brocksopp}, {McGowan}, {Krimm}, {Godet},
  {Roming}, {Mason}, {Gehrels}, {Still}, {Page}, {Moretti}, {Shrader},
  {Campana}, \& {Kennea}}]{2006MNRAS.365.1203B}
{Brocksopp}, C., {McGowan}, K.~E., {Krimm}, H., {et~al.} 2006, \mnras, 365,
  1203

\bibitem[{{Chakravorty} {et~al.}(2016){Chakravorty}, {Petrucci}, {Ferreira},
  {Henri}, {Belmont}, {Clavel}, {Corbel}, {Rodriguez}, {Coriat}, {Drappeau}, \&
  {Malzac}}]{2016A&A...589A.119C}
{Chakravorty}, S., {Petrucci}, P.~O., {Ferreira}, J., {et~al.} 2016, \aap, 589,
  A119

\bibitem[{{Dickey} \& {Lockman}(1990)}]{1990ARA&A..28..215D}
{Dickey}, J.~M., \& {Lockman}, F.~J. 1990, \araa, 28, 215

\bibitem[{{Done} {et~al.}(2018){Done}, {Tomaru}, \&
  {Takahashi}}]{2018MNRAS.473..838D}
{Done}, C., {Tomaru}, R., \& {Takahashi}, T. 2018, \mnras, 473, 838

\bibitem[{{Fukumura} {et~al.}(2017){Fukumura}, {Kazanas}, {Shrader}, {Behar},
  {Tombesi}, \& {Contopoulos}}]{2017NatAs...1E..62F}
{Fukumura}, K., {Kazanas}, D., {Shrader}, C., {et~al.} 2017, Nature Astronomy,
  1, 0062

\bibitem[{{Ganguly} \& {Proga}(2020)}]{2020ApJ...890...54G}
{Ganguly}, S., \& {Proga}, D. 2020, \apj, 890, 54

\bibitem[{{Gehrels} {et~al.}(2004){Gehrels}, {Chincarini}, {Giommi}, {Mason},
  {Nousek}, {Wells}, {White}, {Barthelmy}, {Burrows}, {Cominsky}, {Hurley},
  {Marshall}, {M{\'e}sz{\'a}ros}, {Roming}, {Angelini}, {Barbier}, {Belloni},
  {Campana}, {Caraveo}, {Chester}, {Citterio}, {Cline}, {Cropper}, {Cummings},
  {Dean}, {Feigelson}, {Fenimore}, {Frail}, {Fruchter}, {Garmire}, {Gendreau},
  {Ghisellini}, {Greiner}, {Hill}, {Hunsberger}, {Krimm}, {Kulkarni}, {Kumar},
  {Lebrun}, {Lloyd-Ronning}, {Markwardt}, {Mattson}, {Mushotzky}, {Norris},
  {Osborne}, {Paczynski}, {Palmer}, {Park}, {Parsons}, {Paul}, {Rees},
  {Reynolds}, {Rhoads}, {Sasseen}, {Schaefer}, {Short}, {Smale}, {Smith},
  {Stella}, {Tagliaferri}, {Takahashi}, {Tashiro}, {Townsley}, {Tueller},
  {Turner}, {Vietri}, {Voges}, {Ward}, {Willingale}, {Zerbi}, \&
  {Zhang}}]{2004ApJ...611.1005G}
{Gehrels}, N., {Chincarini}, G., {Giommi}, P., {et~al.} 2004, \apj, 611, 1005

\bibitem[{{Greene} {et~al.}(2001){Greene}, {Bailyn}, \&
  {Orosz}}]{2001ApJ...554.1290G}
{Greene}, J., {Bailyn}, C.~D., \& {Orosz}, J.~A. 2001, \apj, 554, 1290

\bibitem[{{Harmon} {et~al.}(1995){Harmon}, {Wilson}, {Zhang}, {Paciesas},
  {Fishman}, {Hjellming}, {Rupen}, {Scott}, {Briggs}, \&
  {Rubin}}]{1995Natur.374..703H}
{Harmon}, B.~A., {Wilson}, C.~A., {Zhang}, S.~N., {et~al.} 1995, \nat, 374, 703

\bibitem[{{Higginbottom} {et~al.}(2017){Higginbottom}, {Proga}, {Knigge}, \&
  {Long}}]{2017ApJ...836...42H}
{Higginbottom}, N., {Proga}, D., {Knigge}, C., \& {Long}, K.~S. 2017, \apj,
  836, 42

\bibitem[{{Holczer} {et~al.}(2007){Holczer}, {Behar}, \&
  {Kaspi}}]{2007ApJ...663..799H}
{Holczer}, T., {Behar}, E., \& {Kaspi}, S. 2007, \apj, 663, 799

\bibitem[{{Kallman} {et~al.}(2009){Kallman}, {Bautista}, {Goriely}, {Mendoza},
  {Miller}, {Palmeri}, {Quinet}, \& {Raymond}}]{2009ApJ...701..865K}
{Kallman}, T.~R., {Bautista}, M.~A., {Goriely}, S., {et~al.} 2009, \apj, 701,
  865

\bibitem[{{Kammoun} {et~al.}(2019){Kammoun}, {Miller}, {Zoghbi}, {Oh}, {Koss},
  {Mushotzky}, {Brenneman}, {Brand t}, {Proga}, {Lohfink}, {Kaastra}, {Barret},
  {Behar}, \& {Stern}}]{2019ApJ...877..102K}
{Kammoun}, E.~S., {Miller}, J.~M., {Zoghbi}, A., {et~al.} 2019, \apj, 877, 102

\bibitem[{{King} {et~al.}(2015){King}, {Miller}, {Raymond}, {Reynolds}, \&
  {Morningstar}}]{2015ApJ...813L..37K}
{King}, A.~L., {Miller}, J.~M., {Raymond}, J., {Reynolds}, M.~T., \&
  {Morningstar}, W. 2015, \apjl, 813, L37

\bibitem[{{King} {et~al.}(2012){King}, {Miller}, {Raymond}, {Fabian},
  {Reynolds}, {Kallman}, {Maitra}, {Cackett}, \& {Rupen}}]{2012ApJ...746L..20K}
{King}, A.~L., {Miller}, J.~M., {Raymond}, J., {et~al.} 2012, \apjl, 746, L20

\bibitem[{{Luketic} {et~al.}(2010){Luketic}, {Proga}, {Kallman}, {Raymond}, \&
  {Miller}}]{2010ApJ...719..515L}
{Luketic}, S., {Proga}, D., {Kallman}, T.~R., {Raymond}, J.~C., \& {Miller},
  J.~M. 2010, \apj, 719, 515

\bibitem[{{Markwardt} \& {Swank}(2005)}]{2005ATel..414....1M}
{Markwardt}, C.~B., \& {Swank}, J.~H. 2005, The Astronomer's Telegram, 414

\bibitem[{{Marshall} {et~al.}(2002){Marshall}, {Canizares}, \&
  {Schulz}}]{2002ApJ...564..941M}
{Marshall}, H.~L., {Canizares}, C.~R., \& {Schulz}, N.~S. 2002, \apj, 564, 941

\bibitem[{{Miller} {et~al.}(2015){Miller}, {Fabian}, {Kaastra}, {Kallman},
  {King}, {Proga}, {Raymond}, \& {Reynolds}}]{2015ApJ...814...87M}
{Miller}, J.~M., {Fabian}, A.~C., {Kaastra}, J., {et~al.} 2015, \apj, 814, 87

\bibitem[{{Miller} {et~al.}(2006{\natexlab{a}}){Miller}, {Raymond}, {Fabian},
  {Steeghs}, {Homan}, {Reynolds}, {van der Klis}, \&
  {Wijnands}}]{2006Natur.441..953M}
{Miller}, J.~M., {Raymond}, J., {Fabian}, A., {et~al.} 2006{\natexlab{a}},
  \nat, 441, 953

\bibitem[{{Miller} {et~al.}(2008){Miller}, {Raymond}, {Reynolds}, {Fabian},
  {Kallman}, \& {Homan}}]{2008ApJ...680.1359M}
{Miller}, J.~M., {Raymond}, J., {Reynolds}, C.~S., {et~al.} 2008, \apj, 680,
  1359

\bibitem[{{Miller} {et~al.}(2006{\natexlab{b}}){Miller}, {Raymond}, {Homan},
  {Fabian}, {Steeghs}, {Wijnands}, {Rupen}, {Charles}, {van der Klis}, \&
  {Lewin}}]{2006ApJ...646..394M}
{Miller}, J.~M., {Raymond}, J., {Homan}, J., {et~al.} 2006{\natexlab{b}}, \apj,
  646, 394

\bibitem[{{Miller} {et~al.}(2016){Miller}, {Raymond}, {Fabian}, {Gallo},
  {Kaastra}, {Kallman}, {King}, {Proga}, {Reynolds}, \&
  {Zoghbi}}]{2016ApJ...821L...9M}
{Miller}, J.~M., {Raymond}, J., {Fabian}, A.~C., {et~al.} 2016, \apjl, 821, L9

\bibitem[{{Mitsuda} {et~al.}(1984){Mitsuda}, {Inoue}, {Koyama}, {Makishima},
  {Matsuoka}, {Ogawara}, {Shibazaki}, {Suzuki}, {Tanaka}, \&
  {Hirano}}]{1984PASJ...36..741M}
{Mitsuda}, K., {Inoue}, H., {Koyama}, K., {et~al.} 1984, \pasj, 36, 741

\bibitem[{{Nardini} {et~al.}(2015){Nardini}, {Reeves}, {Gofford}, {Harrison},
  {Risaliti}, {Braito}, {Costa}, {Matzeu}, {Walton}, {Behar}, {Boggs},
  {Christensen}, {Craig}, {Hailey}, {Matt}, {Miller}, {O'Brien}, {Stern},
  {Turner}, \& {Ward}}]{2015Sci...347..860N}
{Nardini}, E., {Reeves}, J.~N., {Gofford}, J., {et~al.} 2015, Science, 347, 860

\bibitem[{{Neilsen} \& {Homan}(2012)}]{2012ApJ...750...27N}
{Neilsen}, J., \& {Homan}, J. 2012, \apj, 750, 27

\bibitem[{{Neilsen} \& {Lee}(2009)}]{2009Natur.458..481N}
{Neilsen}, J., \& {Lee}, J.~C. 2009, \nat, 458, 481

\bibitem[{{Neilsen} {et~al.}(2016){Neilsen}, {Rahoui}, {Homan}, \&
  {Buxton}}]{2016ApJ...822...20N}
{Neilsen}, J., {Rahoui}, F., {Homan}, J., \& {Buxton}, M. 2016, \apj, 822, 20

\bibitem[{{Orosz} \& {Bailyn}(1997)}]{1997ApJ...477..876O}
{Orosz}, J.~A., \& {Bailyn}, C.~D. 1997, \apj, 477, 876

\bibitem[{{Pinto} {et~al.}(2016){Pinto}, {Middleton}, \&
  {Fabian}}]{2016Natur.533...64P}
{Pinto}, C., {Middleton}, M.~J., \& {Fabian}, A.~C. 2016, \nat, 533, 64

\bibitem[{{Ponti} {et~al.}(2012){Ponti}, {Fender}, {Begelman}, {Dunn},
  {Neilsen}, \& {Coriat}}]{2012MNRAS.422L..11P}
{Ponti}, G., {Fender}, R.~P., {Begelman}, M.~C., {et~al.} 2012, \mnras, 422,
  L11

\bibitem[{{Remillard} \& {McClintock}(2006)}]{2006ARA&A..44...49R}
{Remillard}, R.~A., \& {McClintock}, J.~E. 2006, \araa, 44, 49

\bibitem[{{Shakura} \& {Sunyaev}(1973)}]{1973IAUS...55..155S}
{Shakura}, N.~I., \& {Sunyaev}, R.~A. 1973, in IAU Symposium, Vol.~55, X- and
  Gamma-Ray Astronomy, ed. H.~{Bradt} \& R.~{Giacconi}, 155

\bibitem[{{Shields} {et~al.}(1986){Shields}, {McKee}, {Lin}, \&
  {Begelman}}]{1986ApJ...306...90S}
{Shields}, G.~A., {McKee}, C.~F., {Lin}, D.~N.~C., \& {Begelman}, M.~C. 1986,
  \apj, 306, 90

\bibitem[{{Sobczak} {et~al.}(1999){Sobczak}, {McClintock}, {Remillard},
  {Bailyn}, \& {Orosz}}]{1999ApJ...520..776S}
{Sobczak}, G.~J., {McClintock}, J.~E., {Remillard}, R.~A., {Bailyn}, C.~D., \&
  {Orosz}, J.~A. 1999, \apj, 520, 776

\bibitem[{{Strohmayer}(2001)}]{2001ApJ...552L..49S}
{Strohmayer}, T.~E. 2001, \apjl, 552, L49

\bibitem[{{Tashiro} {et~al.}(2018){Tashiro}, {Maejima}, {Toda}, {Kelley},
  {Reichenthal}, {Lobell}, {Petre}, {Guainazzi}, {Costantini}, {Edison},
  {Fujimoto}, {Grim}, {Hayashida}, {den Herder}, {Ishisaki}, {Paltani},
  {Matsushita}, {Mori}, {Sneiderman}, {Takei}, {Terada}, {Tomida}, {Akamatsu},
  {Angelini}, {Arai}, {Awaki}, {Babyk}, {Bamba}, {Barfknecht}, {Barnstable},
  {Bialas}, {Blagojevic}, {Bonafede}, {Brambora}, {Brenneman}, {Brown},
  {Brown}, {Burns}, {Canavan}, {Carnahan}, {Chiao}, {Comber}, {Corrales}, {de
  Vries}, {Dercksen}, {Diaz-Trigo}, {Dillard}, {DiPirro}, {Done}, {Dotani},
  {Ebisawa}, {Eckart}, {Enoto}, {Ezoe}, {Ferrigno}, {Fukazawa}, {Fujita},
  {Furuzawa}, {Gallo}, {Graham}, {Gu}, {Hagino}, {Hamaguchi}, {Hatsukade},
  {Hawes}, {Hayashi}, {Hegarty}, {Hell}, {Hiraga}, {Hodges-Kluck}, {Holland},
  {Hornschemeier}, {Hoshino}, {Ichinohe}, {Iizuka}, {Ishibashi}, {Ishida},
  {Ishikawa}, {Ishimura}, {James}, {Kallman}, {Kara}, {Katsuda}, {Kenyon},
  {Kilbourne}, {Kimball}, {Kitaguti}, {Kitamoto}, {Kobayashi}, {Kohmura},
  {Koyama}, {Kubota}, {Leutenegger}, {Lockard}, {Loewenstein}, {Maeda},
  {Marbley}, {Markevitch}, {Matsumoto}, {Matsuzaki}, {McCammon}, {McNamara},
  {Miko}, {Miller}, {Miller}, {Minesugi}, {Mitsuishi}, {Mizuno}, {Mori},
  {Mukai}, {Murakami}, {Mushotzky}, {Nakajima}, {Nakamura}, {Nakashima},
  {Nakazawa}, {Natsukari}, {Nigo}, {Nishioka}, {Nobukawa}, {Nobukawa}, {Noda},
  {Odaka}, {Ogawa}, {Ohashi}, {Ohno}, {Ohta}, {Okajima}, {Okamoto}, {Onizuka},
  {Ota}, {Ozaki}, {Plucinsky}, {Porter}, {Pottschmidt}, {Sato}, {Sato},
  {Sawada}, {Seta}, {Shelton}, {Shibano}, {Shida}, {Shidatsu}, {Shirron},
  {Simionescu}, {Smith}, {Someya}, {Soong}, {Suagawara}, {Szymkowiak},
  {Takahashi}, {Tamagawa}, {Tamura}, {Tanaka}, {Terashima}, {Tsuboi},
  {Tsujimoto}, {Tsunemi}, {Tsuru}, {Uchida}, {Uchiyama}, {Ueda}, {Uno},
  {Walsh}, {Watanabe}, {Williams}, {Wolfs}, {Wright}, {Yamada}, {Yamaguchi},
  {Yamaoka}, {Yamasaki}, {Yamauchi}, {Yamauchi}, {Yanagase}, {Yaqoob},
  {Yasuda}, {Yoshioka}, {Zabala}, \& {Irina}}]{2018SPIE10699E..22T}
{Tashiro}, M., {Maejima}, H., {Toda}, K., {et~al.} 2018, in Society of
  Photo-Optical Instrumentation Engineers (SPIE) Conference Series, Vol. 10699,
  \procspie, 1069922

\bibitem[{{Tingay} {et~al.}(1995){Tingay}, {Jauncey}, {Preston}, {Reynolds},
  {Meier}, {Murphy}, {Tzioumis}, {McKay}, {Kesteven}, {Lovell},
  {Campbell-Wilson}, {Ellingsen}, {Gough}, {Hunstead}, {Jonos}, {McCulloch},
  {Migenes}, {Quick}, {Sinclair}, \& {Smits}}]{1995Natur.374..141T}
{Tingay}, S.~J., {Jauncey}, D.~L., {Preston}, R.~A., {et~al.} 1995, \nat, 374,
  141

\bibitem[{{Trueba} {et~al.}(2019){Trueba}, {Miller}, {Kaastra}, {Zoghbi},
  {Fabian}, {Kallman}, {Proga}, \& {Raymond}}]{2019ApJ...886..104T}
{Trueba}, N., {Miller}, J.~M., {Kaastra}, J., {et~al.} 2019, \apj, 886, 104

\bibitem[{{Waters} \& {Proga}(2018)}]{2018MNRAS.481.2628W}
{Waters}, T., \& {Proga}, D. 2018, \mnras, 481, 2628

\bibitem[{{Wilms} {et~al.}(2000){Wilms}, {Allen}, \&
  {McCray}}]{2000ApJ...542..914W}
{Wilms}, J., {Allen}, A., \& {McCray}, R. 2000, \apj, 542, 914

\bibitem[{{Woods} {et~al.}(1996){Woods}, {Klein}, {Castor}, {McKee}, \&
  {Bell}}]{1996ApJ...461..767W}
{Woods}, D.~T., {Klein}, R.~I., {Castor}, J.~I., {McKee}, C.~F., \& {Bell},
  J.~B. 1996, \apj, 461, 767

\end{thebibliography}
\bibliographystyle{aasjournal}

\end{document}